\documentclass[fleqn,usenatbib]{mnras}
\usepackage[T1]{fontenc}
\usepackage{ae,aecompl}
\newif\ifpre \pretrue
\usepackage{graphicx}   
\usepackage{amsmath}    
\usepackage{amssymb}    
\usepackage{float}
\usepackage{cancel}
\usepackage{xcolor}

\newcommand{\getlength}[1]{\ifx#1\end \let\next=\relax
            \else\advance\count255 by1 \let\next=\getlength\fi \next}
\newcommand{\ifnularg}[1]{ \count255=0 \getlength#1\end \ifnum\count255=0 }
\newcommand{\ifm}{\makebox{}\ifmmode}
\long\def\ifundefined#1#2#3{\expandafter\ifx\csname
  #1\endcsname\relax#2\else#3\fi}
\newcommand{\beq}   { \begin{eqnarray} }
\newcommand{\eeq}[1]{ \ifnularg{#1} end{eanarray} \else
                      \label{#1}\end{eqnarray}    \fi }
\newcommand{\eeql}   { \end{eqnarray} }
\newcommand{\eeqn}   { \nonumber \end{eqnarray} }
\newcommand{\Frac}[2]{\frac{\displaystyle\strut #1}{\displaystyle\strut #2} }

\newcommand{\Cov}{ \mathop{ \rm Cov }\nolimits }

\newcommand{\ntab}[2]{ \multicolumn{1}{#1}{#2} }

\newcommand{\Number}[1]{\ifnum#1<10\relax0\number#1\else\number#1\fi}
\newcommand{\isodate}{
\count151=\time
\divide\count151 by 60
\count151=\count151
\multiply\count151 by 60
\count152=\time
\advance\count152 by -\count151
\divide\count151 by 60
\count152=\count151
\multiply\count151 by 60
\count153=\time
\advance\count153 by -\count151
\Number{\year}.\Number{\month}.\Number{\day}--\Number{\count152}:\Number{\count153}
}
\definecolor{Dred}{rgb}{0.312,0.070,0.070}
\definecolor{Dblue}{rgb}{0.070,0.070,0.312}
\definecolor{Dgreen}{rgb}{0.070,0.312,0.070}
\definecolor{Db}{rgb}    {0.050,0.0,0.320}

\newcommand{\atca}{\mbox{\sc atca-\small{104}}}
\newcommand{\ceduna}{\mbox{\sc ceduna}}

\newcommand{\mopra}{\mbox{\sc mopra}}

\newcommand{\Gaia}{{\it Gaia}}
\newcommand{\Fermi}{{\it Fermi}}

\newcounter{note}
\let\oldmarginpar\marginpar
\renewcommand\marginpar[1]{\-\oldmarginpar[\raggedleft\footnotesize #1]%
{\raggedright\footnotesize #1}}
\newcommand{\Note}[1]{\Rdb{#1}{\addtocounter{note}{1}%
\marginpar{\small\underline{\Rdb{Corr \arabic{note}}}}}}
\newcommand{\note}[1]{\Rdb{#1}}
\renewcommand{\note}[1]{#1}
\renewcommand{\Note}[1]{#1}
\volume{485}
\pubyear{2019}
\pagerange{88--101}
\ifpre \renewcommand{\note}{\relax} \renewcommand{\Note}{\relax}
\fi

\title[The LBA Calibrator Survey --- LCS2]{The Second LBA Calibrator Survey
of southern compact extragalactic radio sources --- LCS2}
\author[Petrov et al.]{
  \parbox[t]{\textwidth}{
     Leonid Petrov$^{1}$\thanks{E-mail:Leonid.Petrov@nasa.gov},
     Alet de Witt$^{2}$,
     \mbox{Elaine M.~Sadler}$^{3,4}$,
     Chris Phillips$^{4}$, and
     Shinji Horiuchi$^{5}$,
  }
\vspace{1.0ex} \\
$^{1}$NASA Goddard Space Flight Center, 8800 Greenbelt Rd, Greenbelt, MD 20771, USA \\
$^{2}$Hartebeesthoek Radio Astronomy Observatory, P.O.Box 443, Krugersdorp
1740, South Africa \\
$^{3}$Sydney Institute for Astronomy, School of Physics, The University of
Sydney, NSW 2006, Australia \\
$^{4}$CSIRO Astronomy and Space Science, PO Box 76, Epping, NSW 1710,
      Australia \\
$^{5}$CSIRO Astronomy and Space Science, Canberra Deep Space Communication
Complex,
            PO Box 1035, Tuggeranong, ACT 2901, Australia
}

\date{Accepted 2019 January 18. Received 2019 January 15; in original form 2018 December 9}

\pagerange{\pageref{firstpage}--\pageref{lastpage}}

\begin{document}

\maketitle
\label{firstpage}

\begin{abstract}

   We present the second catalogue of accurate positions and correlated flux
densities for 1100 compact extragalactic radio sources that were not observed
before 2008 at high angular resolution. The catalogue spans the declination
range $[-90\degr, -30\degr]$ and was constructed from nineteen 24-hour VLBI
observing sessions with the Australian Long Baseline Array at 8.3~GHz. The
catalogue presents the final part of the program that was started in 2008.
The goals of that campaign were 1)~to extend the number of compact radio 
sources with precise coordinates and measure their correlated flux densities, 
which can be used for phase referencing \note{VLBI and ALMA observations}, geodetic 
VLBI, search for sources with significant offsets with respect to \Gaia\ 
positions, and space navigation; 2)~to extend the complete flux-limited 
sample of compact extragalactic sources to the Southern Hemisphere; and 
3)~to investigate the parsec-scale properties of sources from the 
high-frequency AT20G survey. The median uncertainty of the source positions 
is 3.5~mas. As a result of this VLBI campaign, the number of compact radio 
sources south of declination $-40\degr$ which have measured VLBI correlated flux densities 
and positions known to milliarcsecond accuracy has increased by over 
a factor of 6.

\end{abstract}


   \hphantom{.}
   \vspace{-145ex}\noindent \includegraphics[width=0.999\textwidth]{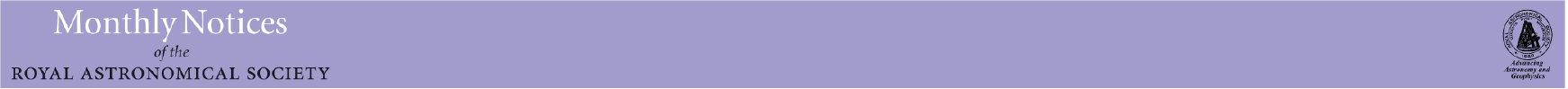}
   \vspace{ 125ex}

\section{Introduction}

   Until recently, the method of very long baseline interferometry
(VLBI) proposed by \citet{r:mat65} was the only way to measure
positions of compact extragalactic radio sources that are almost exclusively
active galactic nuclea (AGNs) with sub-nanoradian accuracy. In 2016,
it has been demonstrated \citep{r:gaia_dr1} that \Gaia\ is able to
get the position accuracy on par with VLBI. However, comparison of VLBI
and \Gaia\ matching sources showed \citep{r:gaia_icrf2,r:gaia1} that there
is a population of sources with statistically significant position offsets.
A more detailed analysis by \citet{r:gaia2}, later extend by \citet{r:gaia4},
revealed that VLBI/\Gaia\ offsets have a preferred direction along the jet
as large as tens milliarcseconds (mas) and the mean of 1--2~mas. They were
interpreted as a manifestation of the contribution of optical jet to the 
centroid position. This allowed \citet{r:gaia3} to make a conclusion that 
VLBI/\Gaia\ differences are due to the fact VLBI and \Gaia\ see different 
part of a source and further improvement in accuracy beyond 1--2~mas level 
will not result in a reconciliation of VLBI and \Gaia\ coordinates of 
active galaxies. A recent publication of \citet{r:gaia5} that used optical 
colors brought additional compelling evidence that synchrotron emission 
from jets shifts the centroid of optical emission along the jet with respect 
to the VLBI positions associated with the jet base. Moreover, the 
VLBI/\Gaia\ offsets brings an important signal that allows us to make 
an inference about milliarcsecond scale source structure of AGNs that 
currently cannot be observed directly. As a consequence, if we need to 
achieve accuracy better than 1--2~mas, we cannot borrow \Gaia\ positions 
of matching sources, but have to rely on VLBI determination of source 
coordinates for applications that needs high accuracy, such as space 
navigation, 
Earth orientation parameter monitoring, and comparison of positions 
of pulsars determined with VLBI and timing.

   In this context, it becomes increasingly important to have an all-sky,
deep, and precise catalogue of positions of extragalactic sources from
radio observations. The most productive instrument for absolute
radio astrometry is the Very Long Baseline Array (VLBA) 
\citep{r:vlba}. Using the VLBA, one can easily determine positions of 
sources at declinations [-$30\degr$, +$90\degr$]
\citep{r:vcs1,r:vcs3,r:vcs4,r:vcs5,r:vcs6,r:obrs1,r:obrs2,r:astro_vips,
r:vgaps,r:bessel_search,r:v2m,r:vcs-ii,r:vcs9};
with some difficulties positions of sources at declinations
[\mbox{-$45\degr$}, \mbox{-$30\degr$}] \citep{r:vcs2}; but with some 
exceptions one cannot observe sources with declinations below 
\mbox{-$45\degr$}. The sequence of VLBA Calibrator Surveys~1--9 (VCS) 
\citep[e.g.,][and references therein]{r:vcs6} provided a dense grid 
of calibrator sources.

  The lack of a VLBA analogue in the Southern Hemisphere resulted in the
past in a significant hemisphere disparity of the source distribution in
absolute radioastrometry catalogues. To alleviate this problem, we launched
a program for observing radio sources at declinations
[-$90\degr$, -$40\degr$] with the Long Baseline Array (LBA) in 2008. The main
goal of the program was to increase the density of calibrator sources with
positions known at milliarcsecond level in the Southern Hemisphere to make
an analogue of the VCS in the south. Unlike the VCS surveys in the Northern
Hemisphere, we predominately used the AT20G survey catalogue \citep{r:at20g} 
from the Australia Telescope Compact Array (ATCA) observations for drawing 
the candidate list for LBA observations. The AT20G is a blind survey that 
covers the Southern Hemisphere. The central frequency of the survey is 20~GHz,
the beamsize $\sim\!10''$, and the catalogue is complete at a 40~mJy level.

  The results of the first part of this campaign for observing the
brightest sources, the catalogue LCS--1 was published by
\citet{r:lcs1}. Here we present results of the second, final part
of the campaign. In the following sections we describe observations,
data analysis, analysis of reported errors, and provide a brief discussion
of results.

\section{Observations}

\subsection{Network}

\begin{figure}
    \includegraphics[width=0.48\textwidth]{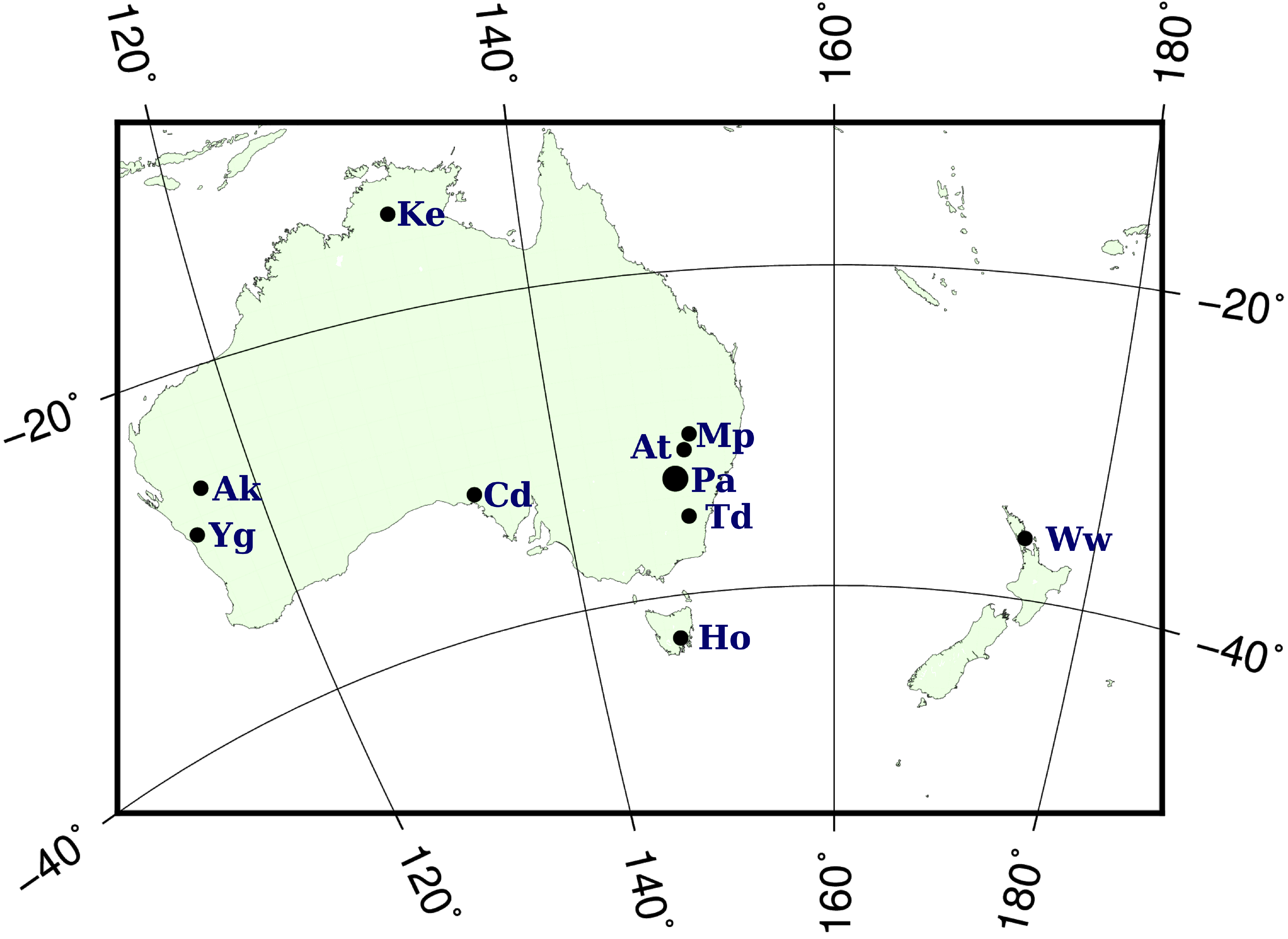}
    \caption{The LBA stations network. Station Hh ({\sc hartrao}), 60~km
             north-west of Johannesburg, South Africa, is not shown.}
    \label{f:network}
\end{figure}

   The observing network includes 11 stations listed in Table~\ref{t:lba}, 
although only a subset of stations participated at any given observing 
session. The list of VLBI experiments, observation dates, and the 
participating network is shown in Table~\ref{t:lcs_exp}. The network, 
except station {\sc hartrao} is shown in Figure~\ref{f:network}.
Station {\sc askap} participated in three experiments, station
{\sc tidbinbilla} observed only during 4--8 hours intervals. The 64~m
station {\sc parkes} was scheduled in every experiment and in every scan
of target sources since it is the most sensitive antenna of the network
and therefore, the sensitivity at baselines with {\sc parkes} is the
highest.

\begin{table}
    \caption{The LBA network. The typical system equivalent flux
             density (SEFD) at 8.3~GHz at elevation angles $> 45\degr$
             achieved in LCS experiments is shown in the last column.}
    \begin{tabular}{l @{\hspace{1.0em}} l @{\hspace{1.0em}} l
@{\hspace{1.0em}} r @{\hspace{1.0em}} l @{\hspace{1.0em}} r}
       \hline
       Code & Name   &  \ntab{c}{$\phi_{gc}$}   &
                        \ntab{c}{$\lambda$}     & Diam & SEFD \\
       \hline
       Ak & \sc askap       & -$26\degr.53$  & $116\degr.63$ &            12 m
& 8300 Jy  \\
       At & \sc atca        & -$30\degr.15$  & $149\degr.57$ & $5 \times 22$ m
&  140 Jy  \\
       Cd & \sc ceduna      & -$31\degr.70$  & $133\degr.81$ &            32 m
&  600 Jy  \\
       Ha & \sc hartrao     & -$25\degr.74$  & $ 27\degr.69$ &            26 m
& 1200 Jy  \\
       Ho & \sc hobart26    & -$42\degr.62$  & $147\degr.44$ &            26 m
&  850 Jy  \\
       Ke & \sc kath12m     & -$14\degr.28$  & $132\degr.15$ &            12 m
& 3000 Jy  \\
       Mp & \sc mopra       & -$31\degr.10$  & $149\degr.10$ &            22 m
&  400 Jy  \\
       Pa & \sc parkes      & -$32\degr.82$  & $148\degr.26$ &            64 m
&   50 Jy  \\
       Td & \sc tidbinbilla & -$35\degr.22$  & $148\degr.98$ &            34 m
&  120 Jy  \\
       Yg & \sc yarra12m    & -$28\degr.88$  & $115\degr.35$ &            12 m
& 3000 Jy  \\
       Ww & \sc wark12m     & -$36\degr.25$  & $174\degr.66$ &            12 m
& 3000 Jy  \\
       \hline
    \end{tabular}
    \label{t:lba}
\end{table}

  Stations {\sc atca, ceduna} and {\sc mopra} were equipped with the LBA VLBI
backend consisting of an Australia Telescope National Facility (ATNF) Data
Acquisition System (DAS) with an Longe Baseline Array data recorder (LBADR)
\citep{r:lbdar}. The ATNF DAS 
only allows two simultaneous intermediate frequencies (IFs): either 2 
frequencies or 2 polarizations. For each of these IFs the input 64~MHz analog 
IF is digitally filtered to 2~contiguous 16~MHz bands. Stations {\sc atca} and 
{\sc mopra} were equipped with two LBDAR recorders, however because of hardware 
limitations, additional recorders could not be used for expanding frequency 
coverage, but could be used for recording both polarizations. Thus, the 
stations equipped with the ATNF backend could record two bands 32~MHz wide. 
This imposed a limitation on the frequency setup: spreading the frequencies 
too narrow would result in a degradation  of group delay accuracy and 
spreading the frequencies too wide would results in group delay ambiguities 
with very narrow group delay ambiguities spacings. Our choice was to spread 
32~MHz sub-bands over 256~MHz band that allowed us to determine group 
delay with uncertainly 123~ps when the signal to noise ratio is 10 and 
with ambiguity of 3.9125~ns.

\begin{figure}
    \centerline{\includegraphics[width=0.44\textwidth]{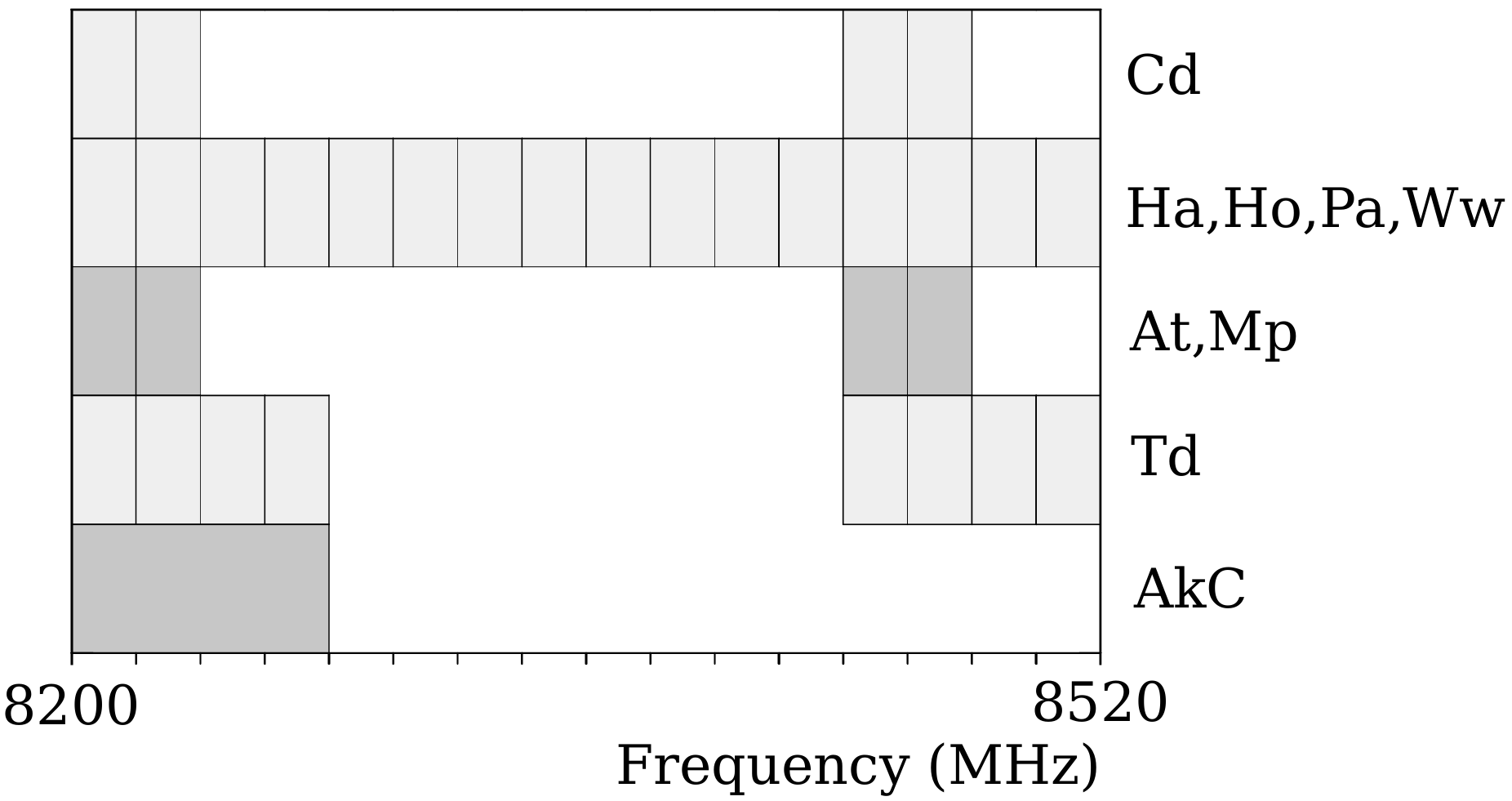}}
    \caption{The frequency allocation in v271i experiment. The channel
             width is 16~MHz for all stations, except Ak, which has
             the channel width 64~MHz. Single polarization channels are
             shown with light-gray color and dual polarization channels
             are shown with dark-Gray color.
             }
    \label{f:v271i_freq_setup}
\end{figure}

  Other stations were equipped with Mark-5 data acquisition terminals.
Station {\sc ceduna} was upgraded from the ATNF backend to Mark-5 in 2015 and 
used Mark-5 in the last three observing sessions. Stations equipped with 
Mark-5 recorded 256~MHz bandwidth, except {\sc tidbinbilla} that prior 2016 
was able to record only 128~MHz, and station {\sc askap} that could record 
a single bandwidth 64~MHz, dual polarization. The stations equipped with 
Mark-5 recorded more 16~MHz wide frequency channels with 320~MHz wide 
spanned bandwidth that partly overlapped with the frequency channels recorded 
by the stations with the ATNF backend. In every experiment from 2 to 5 
different setups were used, and these setups were changing from an experiment 
to experiment. Figure~\ref{f:v271i_freq_setup} as an example shows the frequency
setup of v271i experiment. The versatility of the {\sc difx} correlator
\citep{r:difx2} was exploited to cross-correlate the overlapping regions of 
such experiments. The heterogeneity of the available VLBI hardware made
correlation more difficult but fortunately, did not introduce noticeable
systematic errors in group delay. The most profound effect of this frequency
allocation is ambiguities in group delay at baselines with stations with 
the ATNF backends.

  The telescopes at NASA's Deep Space Network (DNS) located at Tidbinbilla
(Td), near Canberra, participated in the network, when available.
These are Deep Space Station 34 (DSS-34, 34m, for v271h,j,m, and v441a),
DSS-45 (34m, for v271a,b,l), and DSS-43 (70m, for v493a). Their primary mission
is to support communication with spacecrafts but also support VLBI for
celestial reference frame maintenance, navigation, and astronomy for
limited amount of time. For LBA, usually short blocks of 3--5 hour 
were available during time not suited for deep space communication.
Mark-5a system was used to record 8x16MHz channels for this series except
v493a recorded 2x64MHz with LBA-DR. System temperatures at single-band 8GHz
mode the are 22K, 20K, and 12K for DSS-45, DSS-34, and DSS-43,
respectively \citet{r:dsn}. DSS-45 has been decommissioned in November 2016
after operation for 34 years.

\begin{table}
   \caption{List of the LBA Calibrator Survey experiments.
            The first column shows the campaign segment, the second
            and third show the observing session and experiment ID,
            and the last segment shows the network of participating
            stations.}
   \begin{tabular}{ @{\hspace{-0.0em}} l @{\hspace{1.0em}} l
                    @{\hspace{1.0em}} l @{\hspace{1.0em}} l}
      \hline
      LCS--1 & 20080205\_r  & v254b & At-Cd-Ho-Mp-Pa    \\
      LCS--1 & 20080810\_r  & v271a & At-Cd-Ho-Mp-Pa-Td \\
      LCS--1 & 20081128\_r  & v271b & At-Cd-Ho-Mp-Pa-Td \\
      LCS--1 & 20090704\_r  & v271c & At-Cd-Ho-Mp-Pa \\
      \hline
      LCS--2 & 20091212\_r  & v271d & At-Cd-Ho-Mp-Pa \\
      LCS--2 & 20100311\_r  & v271e & At-Cd-Ho-Mp-Pa \\
      LCS--2 & 20100725\_p  & v271f & At-Cd-Ho-Mp-Pa \\
      LCS--2 & 20101029\_p  & v271g & At-Cd-Mp-Pa \\
      LCS--2 & 20110402\_p  & v271h & At-Cd-Ho-Hh-Ww-Td \\
      LCS--2 & 20110723\_p  & v271i & Ak-At-Cd-Ho-Hh-Mp-Pa-Td-Ww \\
      LCS--2 & 20111111\_p  & v271j & At-Cd-Ho-Hh-Mp-Td \\
      LCS--2 & 20111112\_p  & v441a & At-Cd-Ho-Hh-Mp-Td \\
      LCS--2 & 20120428\_p  & v271k & At-Cd-Ho-Hh-Mp-Pa-Ww-Yg \\
      LCS--2 & 20130315\_p  & v271l & Ak-At-Cd-Ho-Hh-Mp-Pa-Ww-Td \\
      LCS--2 & 20130615\_p  & v271m & At-Cd-Ho-Hh-Mp-Pa-Ww-Td \\
      LCS--2 & 20140603\_p  & v493a & At-Cd-Ho-Hh-Mp-Pa-Td \\
      LCS--2 & 20150407\_p  & v271n & At-Cd-Ho-Hh-Ke-Pa-Ww-Yg \\
      LCS--2 & 20150929\_q  & v271o & Ak-At-Cd-Ho-Hh-Ke-Pa-Ww-Yg \\
      LCS--2 & 20160628\_q  & v493c & Ak-At-Cd-Ke-Mp-Pa-Yg \\
      \hline
   \end{tabular}
   \label{t:lcs_exp}
\end{table}

\begin{figure}
    \includegraphics[width=0.48\textwidth]{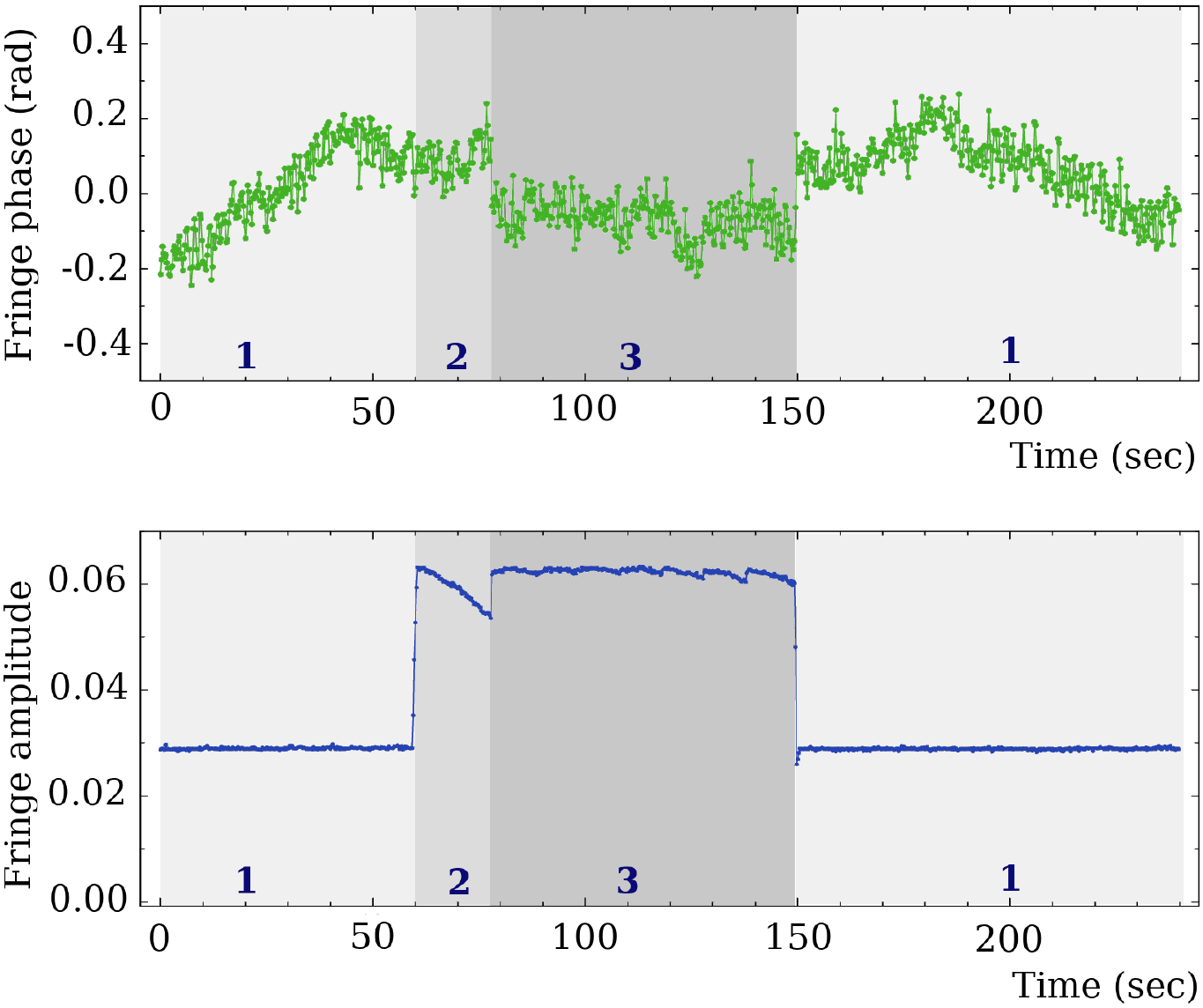}
    \caption{The fringe plot at {\sc atca/parkes}, source {\sf J2225-0457}
             during vt10k test experiment. The upper plot shows fringe phase,
             the lower plot shows fringe amplitude. Light-gray area ``1''
             denotes the interval when single ATCA station records.
             The dark-gray area ``3'' denotes the interval when the
             phased-array records. The medium-gray area ``2'' denotes
             the intermediate interval.}
    \label{f:frplot}
\end{figure}

  The Australia Telescope Compact Array (ATCA) consists of six 22~m antennas.
Five of them can be phased up and record the signal from each individual
telescope as a single element of the VLBI network. The position  of the 
ATCA phase center can be set to any of the antenna positions. However, we 
exercised caution in using the phased ATCA since attempts to use the phased 
Westerbork array for astrometry revealed significant phase fluctuations 
in the past which rendered it highly problematic for precise astrometry 
(Sergei Pogrebenko, private communication, 2010). Therefore, we investigated 
the performance of the phase ATCA in a special 4~hour long test experiment 
that we ran on May 08, 2010. Stations {\sc atca, ceduna, hobart26, mopra}, 
and {\sc parkes} recorded the same frequency setup as in the LCS experiments. 
For the first 60 seconds of a 4 minute long scan the ATCA recorded signal 
from the single antenna at pad with ID W104 (see LCS1 paper for the 
nomenclature of ATCA pads), then it switched to the phased array with the 
phase center at the same pad and recorded for a further 90~second.
Finally, ATCA switched back to recording the signal from a single station.
In total, 232 scans of strong sources were recorded. The typical plots of
the normalized uncalibrated fringe amplitude and fringe phase as
a function of time within a scan are shown in Figure ~\ref{f:frplot}.

  We see that for 18 seconds after switching to the phased-up mode the fringe
amplitude is steadily drops by 15\% and then suddenly returns back and stays
stable within 2\%. We consider this as transitional interval. The fringe phase
does not show any change greater 0.01 rad just after switching back to the 
phased mode, but shows a sudden change in a range of 0.1--0.2~rad after the 
end of the transitional interval and immediately after switching from the 
phased to the single antenna record mode.

  We computed average fringe phases, phase delay rates, group delays, and
group delay rates by running the fringe fitting algorithm trough the same data
three times. During the first processing run we masked out single antenna
recording mode and the first 18~s of the phased recording mode keeping
72~s long data in each scan when ATCA recorded in the phased mode. During
the second processing run we masked out the data when ATCA recorded in the
phased mode. During the third run we processed first 60~s and last 90~s of
each scan when ATCA recorded in the single antenna mode. We referred group
delay and fringe phases to the same common epoch within a scan and formed
their differences.

\begin{figure}
    \includegraphics[width=0.48\textwidth]{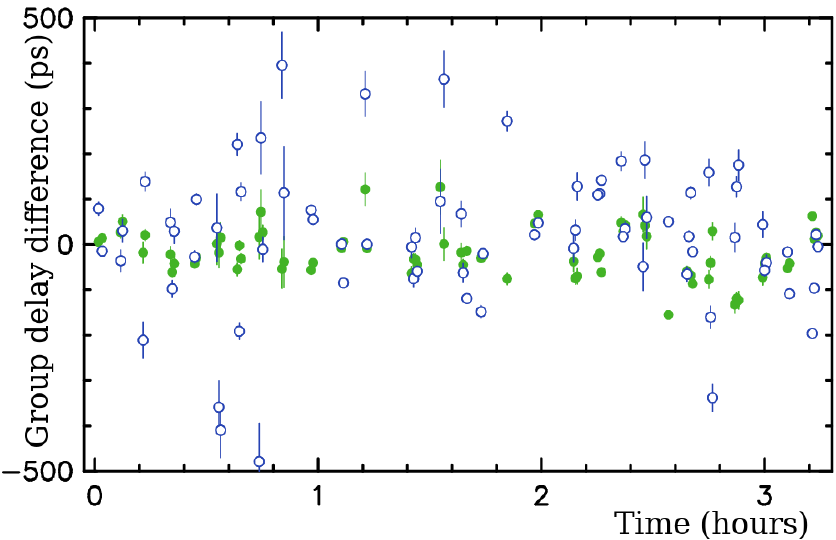}
    \caption{Differences in group delays from the same observations in
             test vt10k experiment. The solid green circles show
             the differences in group delay between ATCA phased-array
             and ATCA-single stations. The wrms of the differences is 38~ps.
             For comparison, the whole blue circles show the differences
             in group delay from first 60 sec and last 90 seconds of
             the integration interval.}
    \label{f:atca_grdel}
\end{figure}

\begin{figure}
    \includegraphics[width=0.48\textwidth]{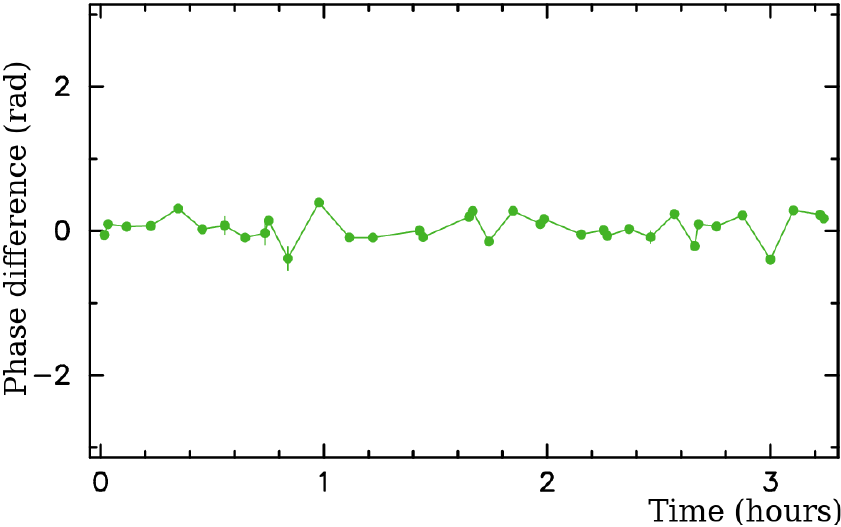}
    \caption{Differences in fringe phase delays between ATCA phased-array
             and ATCA-single recording from the same observations in
             test vt10k experiment. The wrms of the differences ins 0.12~rad.}
    \label{f:atca_phase}
\end{figure}

  The differences in group delay between phased and single antenna
recording modes at different baselines with ATCA are shown on
Figure~\ref{f:atca_grdel} with green color. The weighted root mean square
(wrms) of the differences is 38~ps. For comparison, the differences in group
delays computed using the first 60 seconds and last 90~seconds of a 4~minute
long scan recorded at ATCA in the single antenna mode and referred to the
same middle epoch are shown with blue color. The wrms of these differences
is 59~ps. The differences in fringe phase between recording at ATCA with
phased model and single antenna mode are shown in Figure~\ref{f:atca_phase}.
The wrms of phase differences is 0.12~rad.

  We analyzed the dependence of differences versus elevation, azimuth and
parallactic angle, but found no pattern. The uncalibrated averaged fringe
amplitude at baselines to the ATCA data recorded as a phased array is a factor
of~2.27 greater than the uncalibrated fringe amplitude with ATCA data
recorded as a single antenna, which is within 2\% of $\sqrt{5}$.

  We found that phasing ATCA up does not introduce noticeable systematic
errors in group delay and fringe phases. The differences in group delays
is a factor of 1.5 less than the difference in group delay computed from
two subset of data separated by 90~s. The differences in phases is the random
noise with wrms is 0.12~rad, which corresponds to 0.6~mm. Therefore, we
concluded that using of phased ATCA as an element of the VLBI network does
not introduce systematic errors, but improves sensitivity of ATCA by a factor
of 2.27. This was the first use of a phased array as an element of a VLBI
network for absolute astrometry.

\subsection{Source selection}

   We selected for observations as target sources the objects that had 
previously been detected with single dish observations or with connected 
element interferometers with baselines 0.1--5~km. The input catalogues 
provided estimates of flux density at angular resolutions of 1--$100''$. 
The response of an interferometer to an extended source depends on its 
compactness and the size of the interferometer. The baseline projection 
lengths of the LCS network vary in a range of 5--300~M$\lambda$. That 
means the interferometer will be sensitive for emission from the compact 
components of milliarcsecond size. The response to extended emission with 
a size more than 1~mas at the longest baselines and 50~mas at the shortest 
baselines will be attenuated, and the interferometer will not detect 
signal from emission with size a more than a factor 2--5 beyond that level.

   In order to maximize the number of detected sources, we have to
select the targets with the highest compactness: the ratio of the correlated
flux density at 5--300~M$\lambda$ to the total flux density. As a marker
of high compactness, we initially used spectral index defined as
$S \sim f^{+\alpha}$, where $f$ is the frequency. As a result of synchrotron
self-absorption, the emission from the optically thick jet base that
is morphologically referred to as the core of an AGN, has flat ($\alpha
\approx 0$) or inverted spectrum ($\alpha > 0$). The optically thick
emission from an extended jet and extended radio-lobes that are a result
of interaction of the jet with the surrounding interstellar medium usually 
has steep spectrum ($\alpha < 1$). Therefore, one can expect the sources with
flat spectrum, on average, will have a higher compactness, which has been
confirmed with observations \citep[e.g.,][]{r:vcs1}.

  Our source selection strategy gradually evolved for the course of the 8-yr 
long campaign, but all the time it was focused on selecting the sources with 
brightest correlated flux density. In the first three experiments we
selected sources with spectral index $>-0.50$ from the quarter-Jansky
survey \citep{r:qJy} brighter than 200~mJy. In following experiments we
used several criteria for selecting the targets. In experiments v271c--v271m 
we selected the candidate sources brighter than 150~mJy with spectral index 
$>-0.55$ from the AT20G catalogue. In addition to that, 
we selected sources brighter 180~mJy and spectral index $>-0.55$ from the 
PMN catalogue \citep{r:pmn1,r:pmn2,r:pmn3,r:pmn4,r:pmn5,r:pmn6,r:pmn7}.
The PMN catalogue was derived from processing single-dish observations
with {\sc parkes} at 4.85~GHz, and it is complete at least to 50 mJy
at declinations below $-37^\circ$. In v271k--v271m 
experiments we selected the sources brighter than 170~mJy and
spectral index $>-0.55$ from the ATPMN catalogue \citep{r:atpmn}.
The priority was given to sources with declinations $< -40^\circ$, although
a small fraction of sources with declinations in the range
[$-30^\circ, -40^\circ$] were also observed. We should note here that we
selected {\it some} sources from these pools and did not have intention
to select {\it all} the sources.

  However, an approach of selecting flat spectrum sources does not provide
a good prediction for correlated flux density for the sources within
5--$7^\circ$ of the Galactic plane. The Galactic plane is crowded,
and the chance of making an error in cross-matching the sources observed with
instruments at different angular resolutions and poor positional accuracy
is rather high. This will result in a gross mistake in the estimate of the
spectral indices. Also, the density of galactic sources with flat spectrum,
such as supernova remnants and ultra-compact H~II regions is much higher within
the Galactic plane. An attempt to observe flat spectrum sources in the Galactic
plane by cross-matching the MGPS-2 catalogue at 843 MHz \citep{r:mgps2}
with other catalogues resulted in a detection rate of $\sim\! 10\%$.
To overcome this problem, we used another approach to find candidate sources
in the Galactic plane: we analyzed an IR color-color diagram. 
\citet{r:massaro11} noticed that the blazars occupied a special zone in the 
color-color diagram 3.4--4.6~$\mu$m and 4.6--12~$\mu$m. We analyzed this 
dependence ourselves and delineated the zone that encompasses over 85\% 
compact radio-loud AGNs from the cumulative VLBI catalogue Radio Fundamental 
Catalogue (RFC)
(Petrov \& Kovalev, in preparation\footnote{Preview is available at 
\href{http://astrogeo.org/rfc}{http://astrogeo.org/rfc}}). See section 4.2 
in \citet{r:aofus2} for detail. We tried an alternative approach: we selected 
all the sources within
$5^\circ$ of the Galactic plane and declinations below $-40^\circ$ and flux
density $> 50$~mJy and left those that have cross-matches against IR WISE
catalogue \citep{r:wise,r:neowise} within $30''$. Then we threw away 
sources that are beyond the zone of the 3.4--4.6\,$\mu$m and 4.6--12\,$\mu$m
diagrams containing 85\% radio loud AGNs. We observed the brightest
sources from the remaining sample. The detection rate of this sample
was~57\%.

  In addition to these selection methods, we observed in three experiments,
v441a, v493a, and v493c, the flat spectrum sources brighter 10~mJy that were
detected at 5 and 9~GHz by the Australia Telescope Compact Array (ATCA)
within its error ellipse, i.e. 2--$5'$, of unassociated $\gamma$-ray
sources detected with \Fermi\ mission \citep{r:fermi_1FGL} that
we found in a dedicated program \citep{r:aofus1,r:aofus2,r:aofus3} focused in
finding the most plausible radio counterparts of $\gamma$-ray source. Since
radio-loud $\gamma$-ray AGNs tend to be very compact, the presence of
a radio source detected with a connected interferometer within the error
ellipse of a $\gamma$-ray object raises the probability of being detected 
with VLBI. Observing such sources firstly, fits the primary goal of the LCS 
program and secondly, allows us to find associations to \Fermi\ objects 
that previously have been considered unassociated.

\subsection{Scheduling}

  The experiment schedules were generated automatically with the program 
{\sc sur\_sked}\footnote{See \href{http://astrogeo.org/sur_sked/}
{http://astrogeo.org/sur\_sked/}} in a sequence that minimizes the slewing 
time and obeys a number of constraints. Target sources were observed in three 
to four scans for 2 to 4 minutes long each, except weak candidates to \Fermi\
associations that were observed for 5--10~minutes. VLBI experiments had
a nominal duration of 24 hours. During each session, 80--100 target sources
were observed. The minimum gap between consecutive observations of the same
source was set to 2.5~hours. Station {\sc parkes} was required to participate
in each scan, since it is the most sensitive antenna of the array. After
1.5~hours of observing targets sources, a block of calibrator
sources was inserted. These are the sources picked from the pool of known
compact objects stronger 300~mJy. The block consists of 4 sources, with two
of them observed at each station in the elevations in the range of
10--$30^\circ$ (30--$40^\circ$ for {\sc parkes} that has the low elevation 
limit $31^\circ$) and two observed at elevations 55--$85^\circ$. The goal 
of these observations were: 1)~to improve the estimate of the atmosphere path 
delay in zenith direction; 2)~to connect the LCS catalogue to the accumulative
catalogue of compact radio sources; 3)~to use these sources as bandpass
calibrators; and  4)~to use these sources as amplitude calibrators for 
evaluation of gain corrections.

\section{Data analysis}

   The antenna voltage was sampled with 2-bits with an aggregate
bit rate from 256 to 1024 millions samples per second. The data analysis
chain consist of 1)~correlation that is performed at the dedicated
facility, 2)~post-correlation analysis that computes group delays and phase 
delay rates using the spectrum of cross-correlated data; 3)~astrometric 
analysis that computes source positions, and 4)~amplitude analysis that 
either produces source images or estimates of the correlated flux density 
at the specified range of the lengths of projected baselines.

\subsection{Correlation and post-correlation analysis}

  The first four experiments were correlated with the Bonn Mark4 Correlator.
The data from \atca, \ceduna, and \mopra, originally recorded in LBADR format,
were converted to Mark-5b format before correlation. Post-correlation analysis 
of these data was performed at the correlator using software program 
{\sc fourfit}, the baseline-based fringe fit offered within the Haystack 
Observatory Package Software ({\sc hops}) to estimate the residual group 
delay and phase delay rate. More detail about processing these experiments 
can be found in \citet{r:lcs1}.

  The rest of the experiments were correlated with the {\sc difx} software
correlator \citep{r:difx2} at the Curtin University and then by CSIRO.
The output of the {\sc difx} correlator was converted to FITS-IDI format and 
further processed with {\sc pima} VLBI data analysis software \citep{r:vgaps}. 
The correlator provided the time series of the auto- and cross-spectrum 
of the recorded signal with a spectral resolution 0.25~MHz and time 
resolution 0.25~s. Such a choice of correlation parameters allowed us 
to detect sources within several arcminutes of the pointing direction, 
i.e., everywhere within the primary beam of {\sc parkes} radio telescope 
that has full with half maximum (FWHM) at 8.3~GHz close to $2'$.

   The post-correlator analysis chain includes the following steps:

\begin{itemize}
   \item Coarse fringe fitting that is performed using an abridged grid
         of group delays and delay rates without further refinement.
         The goals of this step is to find at each baseline 10--15
         observations with the highest signal to noise ratio (SNR)
         and detect failures at one or more IFs.

   \item Computation of a complex bandpass using the 12 observations
         with the highest SNR. The complex bandpass describes a distortion
         of the phase and amplitude of the recorded signal with respect
         to the signal that reached the antennas. We flagged at this step
         the IFs that either were not recorded or failed. We used
         12 observations for redundancy in order to evaluate the statistics
         of a residual deviation of the phase and amplitude as a function of
         frequency from the ideal after applying the bandpass computed over
         the 12 observations using least squares. Large residuals triggered
         detailed investigation that in a case of a serious hardware
         failure resulted in flagging affected spectral channels.

   \item Fine fringe fitting that is performed with using the
         complex bandpass and the bandpass mask derived in the previous
         step. The preliminary value of the group delay and phase delay
         is found as the maximum element of the two-dimensional Fourier
         transform of the time series of cross-correlation spectrum sampled
         over time and frequency with a step 4 times finer over each dimension
         than the original data. The final value of the group delay and phase
         delay rate is adjusted from phases of the cross-correlation function
         (also known as fringe phases) as small corrections to the preliminary
         values using least squares. Phase residuals of the cross-correlation
         spectrum are analyzed and additive corrections to the a~priori weights
         are computed on the basis of this analysis. The uncertainties of
         estimates of group delays are derived from uncertainties of fringe
         phases and additive weights corrections. The uncertainties of fringe
         phases depend on fringe amplitudes. The explicit expression can be
         found on page 233 of \citet{r:tms}, equation 6.63.

   \item Computation of total group delays and phase delay rates. The
         group delays and phase delay rates derived at the previous
         step are corrections to the a~priori delays and phase delay rates
         used during correlation. The mathematical model of the a~priori group
         delay and phase delay rate used by the correlator is expanded over 
         polynomials of the 5th order at 2~minute long intervals that cover the 
         time range of a VLBI experiment. Using these coefficients, the a~priori 
         group delays and phase delay rates are computed to a common epoch 
         within a scan for the event of arriving the wavefront at a reference 
         station of a baseline. Using these a~priori group delays and phase 
         delay rates, the total group delays for that epoch are formed.
\end{itemize}

\subsection{Astrometric analysis}

  Total group delay is the main observable for astrometric analysis. During
further analysis, the a~priori model of group delay, more sophisticated than
that used for correlation, is computed, and the differences between observed
and theoretical path delays are formed. The partial derivatives of this model
over source coordinates, station positions, the Earth ordination parameters,
atmosphere path delay in zenith direction, and clock function are also 
computed. Then corrections to those parameters are adjusted using least 
squares.

\begin{figure}
    \includegraphics[width=0.48\textwidth]{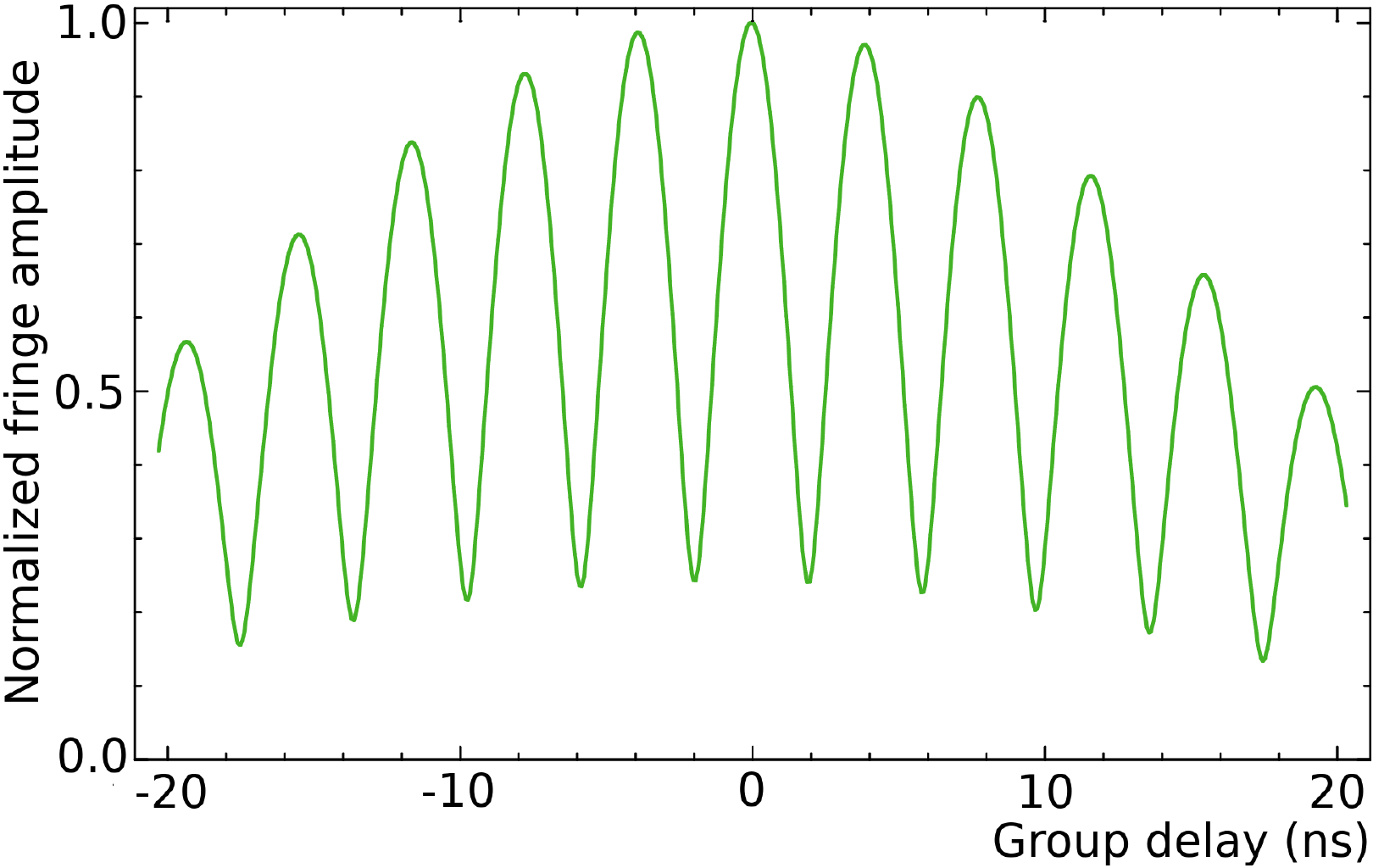}
    \caption{The normalized fringe amplitude as a function of group delay
             between stations that had LBDAR recording system. The fringe
             amplitude is divided by the amplitude at the global maximum.
            }
    \label{f:drf}
\end{figure}

  The frequency setup used for this campaign, selected due to hardware
limitation (see as an example the setup for v271i segment in
Figure~\ref{f:v271i_freq_setup}) posed a challenge in data analysis. The
Fourier transform over frequency over baselines with {\sc atca, ceduna, mopra}
in this example that uses LBDAR data acquisition system has strong secondary
maxima (see Figure~\ref{f:drf}). The amplitude of the 2nd maximum is 0.98, the
third maximum 0.93, and the fourth maximum 0.83 with respect of the global
maximum. Due to the noise in data and \Note{remaining instrumental} phase 
distortion, the fringe fitting process cannot reliably distinguish the 
primary and the secondary maxima, and as a result, group delay is 
determined with the ambiguity of $N \times 1/2.56 \cdot 10^{8} \approx 3.9$~ns, 
where $N$ is a random integer number, typically in a range [-2,2].

  At the first stage of the astrometric data analysis we processed 
the so-called narrow-band group delays derived as an arithmetic mean of
group delays computed over each IF independently and estimated source
positions using least squares. The dataset of narrow-band group delays was
cleaned for outliers during the residual analysis procedure. The narrow-band 
delays do not have ambiguities, but are one order of magnitude less precise 
than group delays computed over the entire band. The estimated parameters 
at this stage are station positions and coordinates of target sources, 
as well as atmospheric path delays in zenith direction and clock function in 
a form of expansion over the B-spline basis. The contribution of the adjusted 
parameters to path delay computed using the narrow-band delays is substituted
to the group delay residuals and then used for initial resolving group delay 
ambiguities. The procedure for group delay ambiguity resolution is described 
in detail in \citet{r:lcs1}.

  After the group delay ambiguities are resolved, the dataset is cleaned for
outliers in group delay. If necessary, the parametric model of clock is refined
for incorporating discontinuities at specified epochs. Initial data weights
were chosen to be reciprocal to the group delay uncertainty $\sigma_g$. Then
the additive baseline-dependent weight corrections $a$ were computed for each
observing session to make the ratio of the weighted sum of residuals be close
to their mathematical expectation. These weights we used in the initial
solution. The weights used in the final solution had a form
\beq
  w = \Frac{1}{k \cdot \sqrt{\sigma_g^2 + a^2 + b^2}},
\eeq{e:e1}
where $k$ is a multiplicative factor and $b$ an additive weight correction
for taking into account mismodeled ionosphere contribution to group delay
(see below). Such a clean dataset of group delays is used in further analysis.

  The final LCS catalogue was derived in a single least square solution using
all dual-band X/S (8.4/2.3 GHz) observations since 1980 through July 2018 
under geodesy and astrometry programs that are publicly available and 19 LCS 
X-band experiments. The estimated parameters are split into three groups:
global parameters that are adjusted for the entire dataset, local parameters
that are specific for a given experiment, and segmented parameters that
are specific for a time interval shorter than the observing session duration.
The global estimated parameters are coordinates of all observed sources,
positions and velocities of all observing stations, harmonic variations of
station positions at annual, semi-annual, diurnal, and semi-diurnal
frequencies, as well as B-spline coefficients that describe discontinuities
and a non-linear motion of station caused by seismic activity. The local
parameters are pole coordinates, UT1, and their first time derivative. The
segmented parameters are clock function for all the stations, except the
reference one, and residual atmosphere path delays in zenith direction.

   For the course of the LCS campaign, a number of target sources were 
observed in follow-up VLBI experiments. We excluded these sources from 
the list of dual-band experiments in our LCS solution. The position of 
target sources were derived using only 8.3~GHz LCS data. Observations of 
these sources were later used for LCS catalogue error evaluation and 
computation of the weight correction factor $k$.

  Since equations of electromagnetic wave propagation are invariant with
respect to a rotation of the celestial coordinate system, as well as
a translation and rotation of the terrestrial coordinate system, the 
system of equations has a rank deficiency and determines only a family 
of solutions. In order to define the solution from that family, we 
applied no-net-rotation constraints for source coordinates requiring 
the new positions of 212 so-called defining sources have no 
net-rotation with with respect to their positions in the ICRF1 
catalogue \citep{r:icrf1}. Similarly, we imposed the no-net-rotation 
and no-net-translation constraints on station positions and velocities.

\subsection{Imaging analysis}

  We derived images of observed sources from one LCS experiment v271e that 
was run on March 2010 and used 5 LBA antennas. A list of 155 sources was 
observed in that experiment, and of them, 122 have been successfully 
detected, including 80 target sources and 42 troposphere calibrators. 
The expected theoretical thermal noise range is between 0.1 and 
0.4~mJy/beam depending on the number of antennas and number of scans 
per source.

\begin{figure*}
   \includegraphics[width=0.325\textwidth]{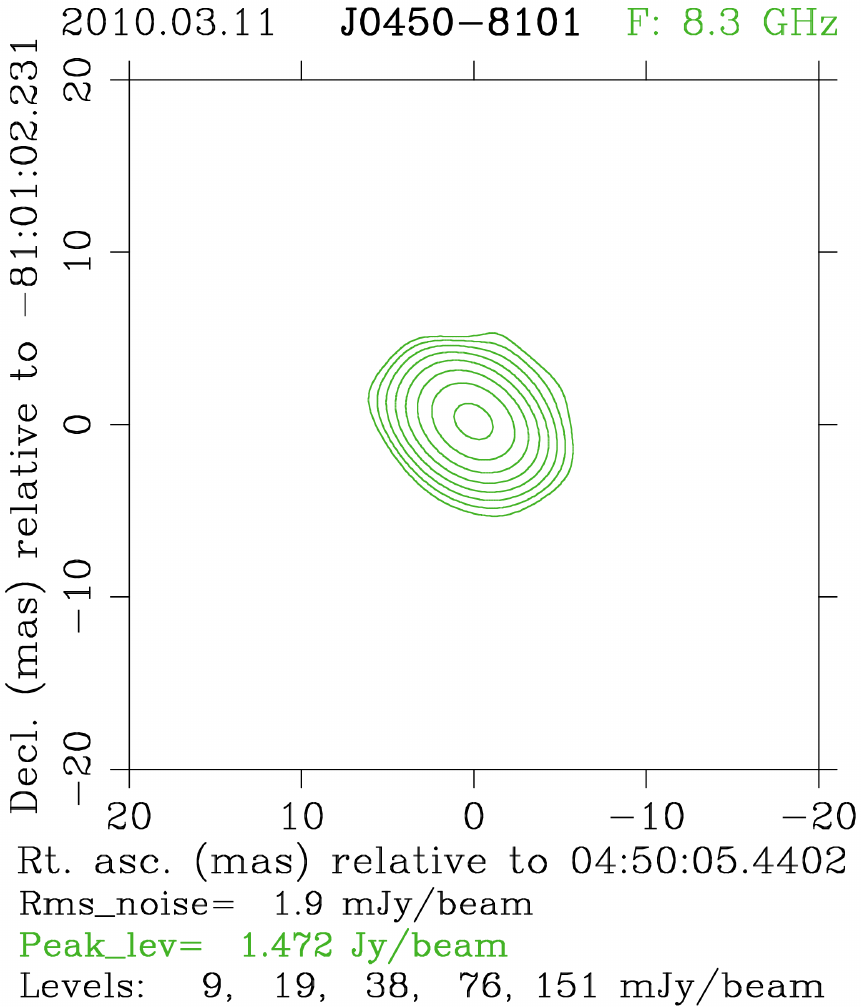} \hspace{0.001\textwidth}
   \includegraphics[width=0.325\textwidth]{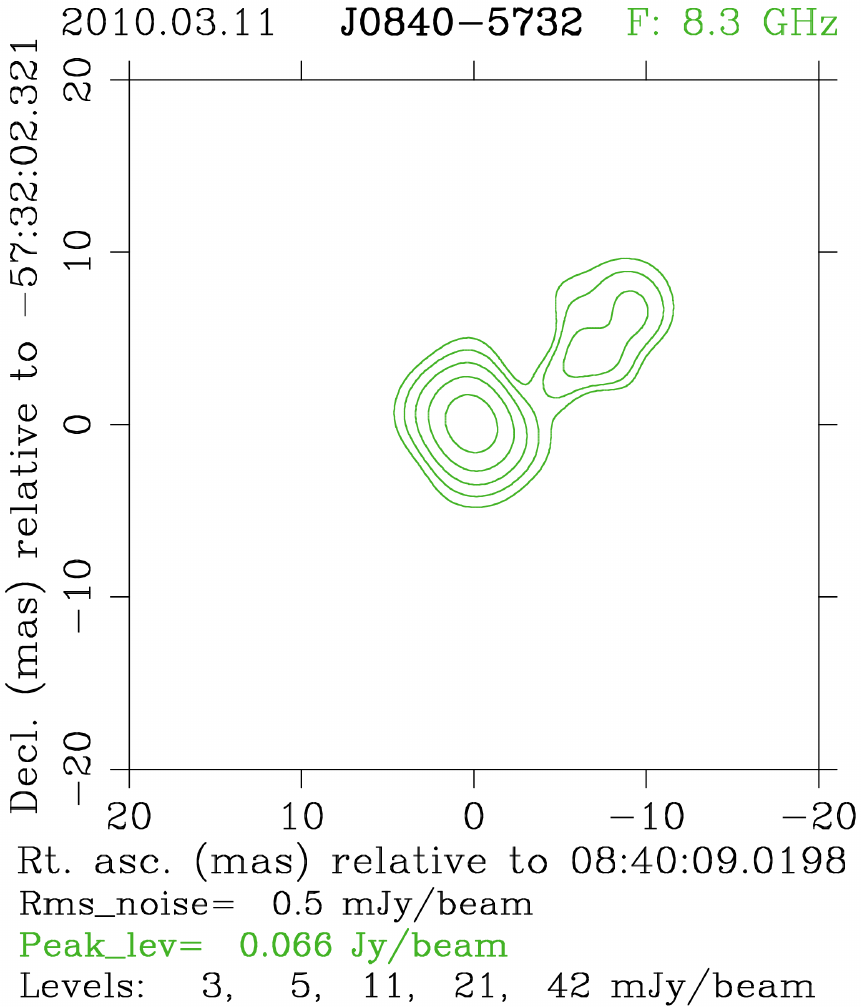} \hspace{0.001\textwidth}
   \includegraphics[width=0.325\textwidth]{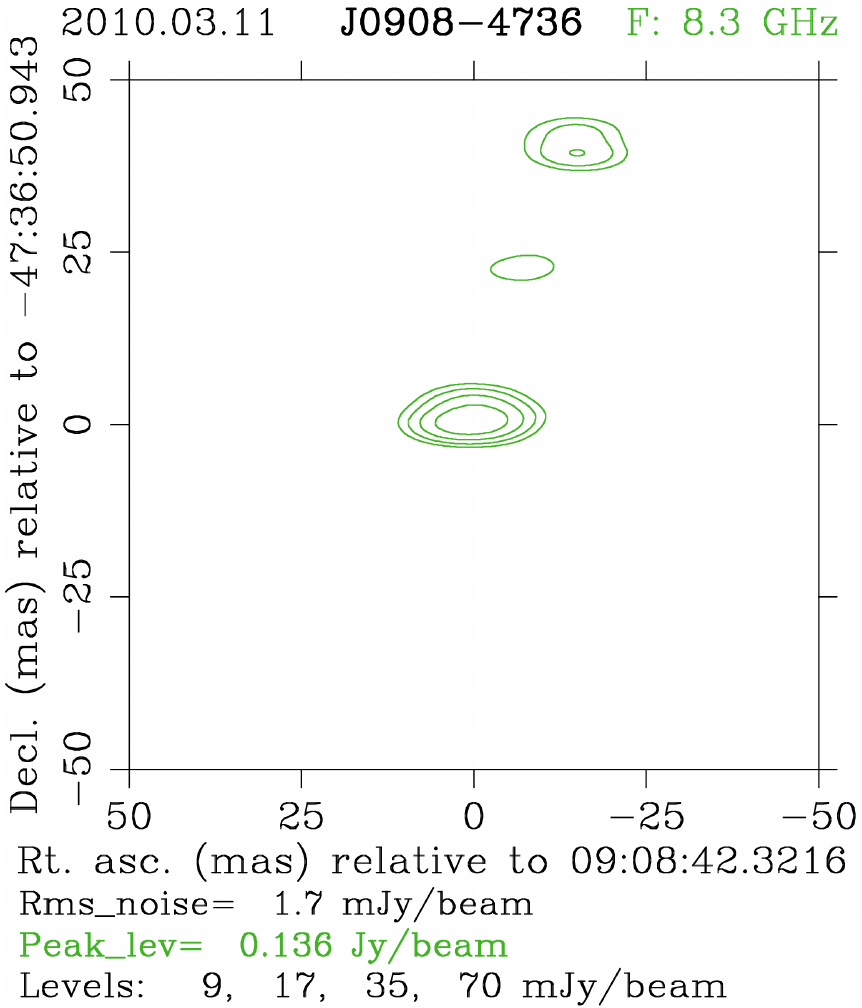}

   \caption{From left to right, contour plots for sources J0450$-$8101,
            J0840$-$5732, and J0908$-$4736 from LCS experiment v271e.
            The declination axis is towards up and the horizontal is 
            towards left.
           }
   \label{f:lcsmaps}
\end{figure*}

  The correlated visibility data were processed using NRAO Astronomical Image 
Processing System ({\sc aips}) software suite of programs \citep{r:aips} 
independently from astrometric analysis. The data were read into AIPS together 
with tables containing system temperature information extracted from observing
logs and antenna gain curves later used to calibrate the visibility amplitudes. 
When system temperature measurements during observations were not available, 
nominal values were used instead. Amplitude gain corrections in the 
cross-correlation spectra due to sampling have also been applied. Thereafter, 
the data inspection, initial editing and fringe fitting were done in the 
traditional manner using {\sc aips}. 

  Fringe fitting in image mode was done independently from fringe fitting
for the astrometry data analysis. We run fringe fitting with {\sc aips} two times. 
The first run of fringe fitting was used before the bulk of the editing to 
find a preliminary approximation for the residual rates and delays. The 
estimates of group delay and phase delay rate were smoothed in time, and 
then used to calibrate the visibilities. The main editing of the visibility 
data was then done using this approximate calibration. Finally, a second 
run of fringe fitting using the edited data was used to refine the rate 
and delay calibration.

  The overall amplitude gains were then further improved by self-calibrating 
five of the most compact and brightest calibrator sources and then using 
a CLEAN image of each of them as a model to determine the antenna gains that 
will make the visibility data for the chosen source/model conform as closely 
as possible to a point source. The derived gain corrections were averaged over 
all  five sources an then applied to the remainder of sources.  

  After further inspection of the quality of the calibrated visibility data, 
the sources were then further self-calibrated in amplitude and phase using 
a CLEAN image of the source itself as a model. We made final CLEAN images 
using a weighting function of the visibilities in between uniform and natural 
weighting and using the square root of the statistical visibility weights. 
It was not possible to conclusively determine the basic source structure 
for a number of sources, and we did not attempt deconvolution or further 
self-calibration for such objects. The \textit{uv}~coverage for the LCS 
experiments is often poor and it is not possible to make satisfactory 
images for every detected source. In total, there were five target sources 
for which we were not able to produce an image using v271e data:
J0413$-$5332, J2103$-$3058, J2107$-$4828, J2239$-$3609 and J2359$-$6054. 
In Figure~\ref{f:lcsmaps} we show representative contour plots from imaging 
results obtained from the v271e session of the LCS.

  In addition to imaging, the correlated flux density and FWHM source size 
were determined by fitting a simple Gaussian model of the emission directly 
to the visibilities by least-squares fitting (using the {\sc aips} task UVFIT). 
The fitted values are listed in Table~\ref{t:lcsfit}. Because of the 
limited, and sometimes highly elongated \textit{uv}~coverage, we did not 
attempt to characterize the source geometry beyond that of a mere estimate 
of the scale of the source structure from a simple circular Gaussian model.
A circular Gaussian model was used, because it has a small number of
parameters. The free parameters were the source position (x,y), the peak
flux density and the FWHM of the Gaussian that fits the source image. 
Visibility plots and CLEAN images (where possible), were examined to 
make sure that a circular Gaussian model is reasonable and that the 
source is not a double for example. 

\begin{table}
   \scalebox{0.86}{
      \begin{tabular}{rrrrr}
          \hline
          \ntab{c}{(1)} & \ntab{c}{(2)} & \ntab{c}{(3)} & \ntab{c}{(4)} & \ntab{c}{(5)} \\
            &  & \ntab{c}{Jy}  & \ntab{c}{mas}  & \ntab{c}{mas}  \\
          \hline
          LCS J0049$-$5738 & 0047$-$579 & 1.411 &  4.91 & -1.00 \\
          LCS J0058$-$5659 & 0056$-$572 & 0.942 &  3.89 & -1.00 \\
          LCS J0109$-$6049 & 0107$-$610 & 0.420 &  8.94 & -1.00 \\
          LCS J0124$-$5113 & 0122$-$514 & 0.171 &  4.10 & -1.00 \\
          LCS J0236$-$6136 & 0235$-$618 & 0.291 &  2.77 & -1.00 \\
          LCS J0314$-$5104 & 0312$-$512 & 0.181 &  2.81 & -1.00 \\
          LCS J0335$-$5430 & 0334$-$546 & 0.560 &  8.24 & -1.00 \\
          \hline
      \end{tabular}
   }\hfill
   \caption{The first 7 rows of the table with results of estimated
            source flux density and the FWHM of the Gaussian fit from the v271e
            observations. The estimated uncertainty on the flux densities
            are 10\%. Fitted FWHM sizes are given with either its
            estimated 1$\sigma$ uncertainty, or the 3$\sigma$ upper limit.
            Columns:
            1)~source J2000 name;
            2)~IVS name;
            3)~flux density;
            4)~FWHM size;
            5)~FWHM size uncertainty (-1.00 if unavailable);
            This table is available in its entirety in machine-readable
            table datafile2 and Virtual Observatory (VO) forms in the online
            journal. A portion is shown here for guidance regarding its form
            and content.
           }
    \label{t:lcsfit}
\end{table}

  For estimating the uncertainty of the FWHM of the Gaussian model, one 
approach would be to just take the statistical uncertainty calculated from 
the model fitting. However, since some residual antenna-dependent calibration
errors are likely, the fitted FWHM size can be strongly correlated with
the antenna amplitude gains, and the previous assumption may be violated.
For this reason, we also estimated the uncertainties of estimated parameters
by redoing the model fitting for all of the sources, but this time with
the antenna amplitude gains added as free parameters. However, because
of the small amount of data, the per-source and per-antenna gain corrections
estimated this way are not always reliable, so we do not use the value of 
the FWHM size, but only the estimate of its uncertainty, which takes into
account the correlation between the antenna gains and the fitted FWHM size.
We found that the statistical uncertainties calculated from the model
fitting are $\sim\! 15$--20\% larger when the antenna amplitude gains
are added as free parameters. 

  To estimate more realistic uncertainties, in particular, to take into
account the contribution of residual mis-calibration, we also used
a Monte Carlo simulation to perturb randomly the antenna gains for
a number of trials and fit the FWHM size. In the absence of reliable estimates
of antenna gains, we just randomly changed them to get a distribution of 
the fitted FWHM size. We assume the antenna gains are accurate at a level 
of 10\%. Therefore, we calculated the uncertainty of the fitted FWHM size from 
the Monte Carlo simulation that is still based on an assumed 10\% uncertainty 
in the antenna gains. The Monte Carlo simulations were carried out using 
10 sources with 12 trials for each source. In each of the trials we varied 
the antenna gains by a random amount, with the mean of the random gain 
variation being 0 and the standard deviation being 10\%. Although the number 
of trials was small, it is enough to determine the scale of this contribution
to the overall uncertainty in the FWHM size, and to show that it is not 
the dominant contribution.

  The estimated uncertainties for the FWHM size from the Monte Carlo
simulations were always  $< a$, the minimum measurable size in the
visibility plane given by \citet{r:L2000}, determined as
\beq
  a < 240 \, \Frac{\sqrt{N/S}}{U},
\eeq{e:a1}
  where $a$ is in units of milliarcsecond, $N$ is the integrated
rms noise of the observations in Janskys and $U$ is the maximum
baseline length in units of M$\lambda$. Expressed in terms
of a uniformly weighted beam size, Equation~\ref{e:a1} can be 
written~as
\beq
  a <  1.8\sqrt{N/S} \times \theta_{\rm beamsize},
\eeq{e:12}
where unites for $a$ and $\theta$ are milliarcseconds.

  Due to the poor \textit{uv}-coverage for these observations, a reliable
estimate of the beamsize could not always be obtained and thus, we
calculated the beamsize as the geometric mean of the beam major and
minor axes of the CLEAN beam. The fitted value of the beamsize for
the v271e observations is $\sim 4$ milliarcseconds. 

From the error calculations described above, the maximum obtained value
of the uncertainty were always $\leq 1/4$ of the beamsize. Due to the
limited \textit{uv}~coverage and the simple approximation of a circular
model, we used 1/4 of the beamsize as a very conservative estimate
of the uncertainty for all of the sources. From the Monte Carlo
simulations the uncertainty on the flux densities were 10\%.

\subsection{Non-imaging analysis}

   As we see in Figure~\ref{f:lcsmaps}, the quality of images is not
great because of scarcity of data and a poor $uv$-coverage. Direct imaging
either produces maps with a dynamic range around 1:100 with a high chance
of an imaging artefact to be unnoticed or, if to pursue elimination of
artefacts aggressively, the images will be close to a point-source or
a single component Gaussian.

   Recognizing these challenges, we processed the entire datasets 
by fitting a simplified source model to calibrated visibilities. 
We limited our analysis to evaluation of the median correlated flux 
density estimates in three ranges of lengths of the baseline projections
onto the plane tangential to the source, without inversion of
calibrated visibility data using the same technique as we used for
processing first 4 LCS experiments \citep{r:lcs1}. A reader is
referred to this publication for detail. Here we outline the procedure.

   At the first step, we analyze system temperatures, remove outliers,
evaluate the radiative atmosphere temperature, compute receiver
temperatures, interpolate them for restoring missing data, and generate
a cleaned dataset of system temperatures. Dividing it by the a~priori
elevation-dependent antenna gain, we get the a~priori system equivalent
flux density (SEFD).

  At the second step, we estimate station-dependent multiplicative gain
corrections to calibrated fringe amplitudes of calibrator sources with
least squares by using a number of sources with known 8~GHz images that
can be found in the Astrogeo VLBI FITS image
database\footnote{Available at \href{http://astrogeo.org/vlbi\_images/}
{http://astrogeo.org/vlbi\_images/}}
that we maintain. During this procedure, we iteratively exclude those
images which resulted in large residuals. Due to variability, the flux
density of some individual sources may raise or decline, but the average
flux density of a source sample is expected to be more stable than
the flux density of individual objects.

  At the third step, we apply adjusted SEFDs and compute the correlated flux
densities of target sources. Then we sort the fringe amplitude over
baseline projection lengths and compute median estimates of the correlated
flux density in three ranges: 0--10~M$\lambda$ ($<360$~km),
10--40~M$\lambda$ (360-1440~km) and 40--300~M$\lambda$ (1400--10800~km).
These parameters characterize the strength of a source, and it has to be
accounted for scheduling the observations. The accuracy of this procedure
is estimated at a level of 20\% judging on residuals of gain adjustments.

  The list of 49915 estimates of correlated flux densities from individual
observations of 1100 target sources and 368 calibrator sources is presented
in the machine-readable table datafile4. The table contains  the following
information: source name, date of observations, baseline name, $u$- and $v$-
projections of the baseline vector, the correlated flux density and their
formal uncertainty, the signal to noise ratio, the instants system
equivalent flux density for this observation, and the observing session
code.

\section{Error analysis}

  Single-band group delays are affected by the contribution of the ionosphere.
Considering the ionosphere as a thin shell at a certain height above the Earth
surface (typically 450 km), the group delay can be expressed as
\beq
  \tau_{\rm iono} = \Frac{\alpha}{f^2_{\mbox{eff}} } \:
                \mbox{TEC} \: \Frac{1}{\cos{\beta}},
\eeq{e:e2}
  where $f_{\mbox{eff}}$ is the effective frequency, $\beta$ is the zenith
angle at the ionosphere piercing point, TEC is the total electron contents
in the zenith direction at the ionosphere piercing point, and  $\alpha$ is
a constant (see \citet{r:sovers98} for detail). We have computed the a~priori
ionosphere contribution to group path delay using TEC maps from analysis
of Global Navigation Satellite System (GNSS) observations. Specifically,
we used CODE TEC time series \citep{r:schaer99}\footnote{Available at
\href{ftp://ftp.aiub.unibe.ch/CODE}{ftp://ftp.aiub.unibe.ch/CODE}} with
a resolution of $5^\circ \times 2.5^\circ \times 1^h$
($5^\circ \times 2.5^\circ \times 2^h$ before December 19, 2013).

  The TEC model from GNSS observation is an approximation, and the accuracy
of a~priori $\tau_{\rm iono}$ from such a model is noticeably lower than the
accuracy of $\tau_{\rm iono}$ computed from the linear combination of group
delays at X and S (or X and C) bands from dual-band observations. The errors
of $\tau_{\rm iono}$ from such observations is at level of several picoseconds
according to \citet{r:hob05}. We consider the contribution of mismodeled
ionospheric path delay as the dominating source of systematic errors and,
therefore, we investigated it in detail.

  We used the global dataset of VLBI observations after July 01, 1998 for
investigation of the residual ionospheric contribution to group delay after
applying the a~priori path delay delay derived from CODE global TEC maps.
For each dual-band observing sessions, we decompose the slant ionospheric
path delay from X/S observations at the product of the path delay in
zenith direction and the mapping function, the ratio of the ionospheric
path delay in a given elevation to the ionospheric path delay in
zenith direction. Then we computed the rms of the total $\tau_{\rm iono}$
in zenith direction from CODE global TEC maps, $\sigma_t$, and the rms of
the differences in $\tau_{\rm iono}$ in zenith direction derived using
the CODE global TEC maps and the dual-band X/S group delays, $\sigma_r$.
The ratio of these two statistics, variance admittance
$A = \frac{\sigma^2_r}{\sigma^2_t}$, is a measure of the model goodness.
Assuming $A$ is stable in time, we can predict, unknown to us, statistics 
of $\sigma^2_r$ for single-band observations using $\sigma^2_t$ that we can
compute from the TEC model. We derived time series of parameter $A$ from
analysis of dual-band VLBI observations after July 01, 1998.

\begin{figure}
    \includegraphics[width=0.48\textwidth]{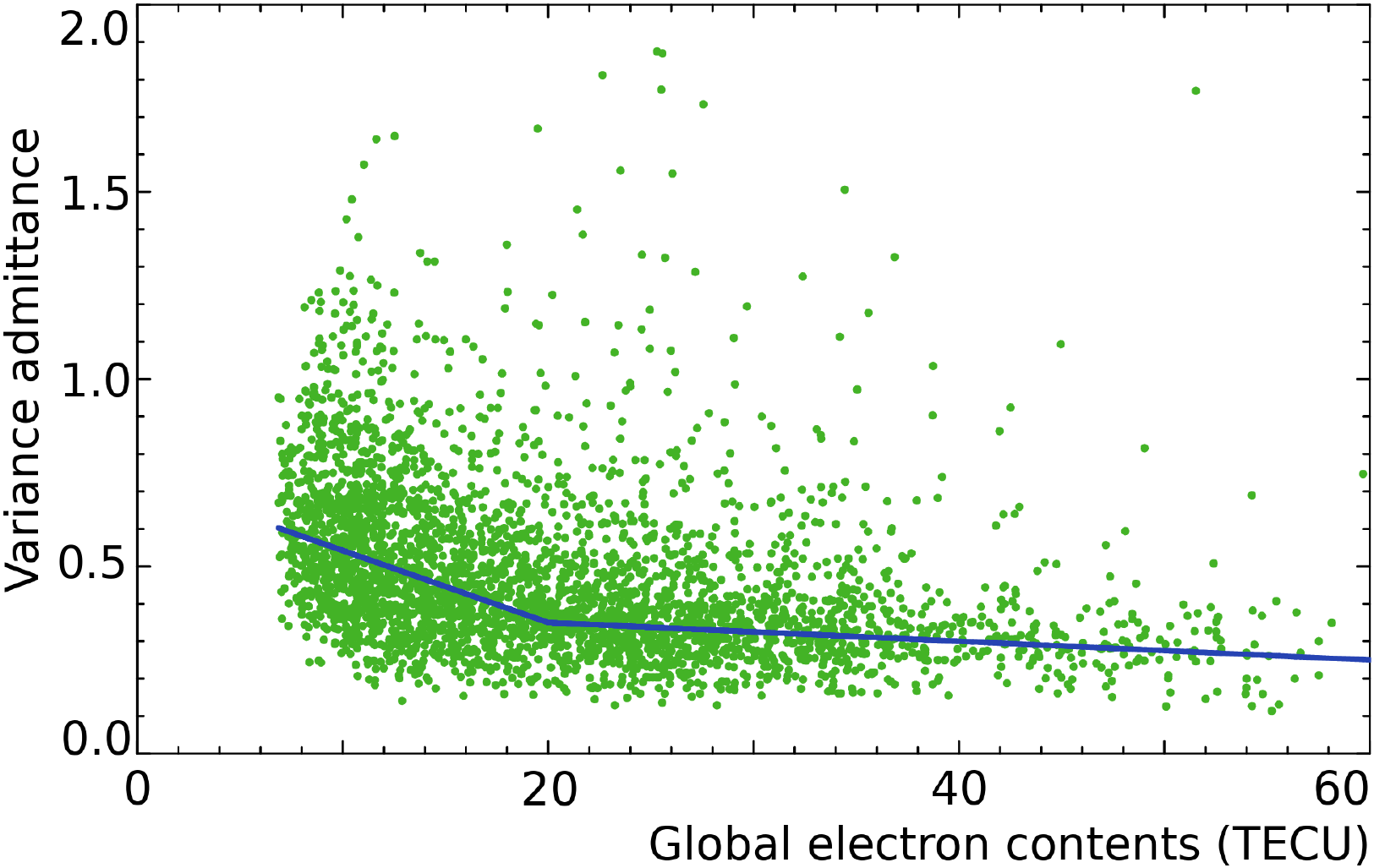}
    \caption{The dependence of the variance admittance factor $A$ on the
             global total electron contents. The so-called TEC units
             ($10^{16}$ electrons over zenith direction) are used for GTEC.}
    \label{f:gtec}
\end{figure}

  Further analysis showed that parameter $A$ is not stable with time. Since 
$A$ is computed as a ratio of variances, we sought an empirical regression 
models where $A$ enters as a multiplicative factor. We computed the global 
total electron content (GTEC) by averaging the TEC over the sphere. 
As it was show by \citet{r:gtec}, such a parameter characterizes the global 
state of the ionosphere. Figure~\ref{f:gtec} shows the dependence of $A$ on 
GTEC. We represent this dependence with a broken linear function with 
A=0.6 at GTEC=7.0, A=0.35 at GTEC=20.0 and A=0.25 at GTEC=60.0.

  Using this dependence, we computed the GTEC for a given experiment, averaged
it over the period of the experiment duration, computed parameter $A$ using 
the linear regression, computed the time series of the ionospheric contribution
from the TEC model for each station of a baseline, and then computed the
variances of the mismodeled contribution of the ionosphere to group delay in
zenith direction for the first and second station of a  baseline, $\Cov_{11}$
and $\Cov_{22}$, as well as their covariances $\Cov_{12}$. Then for each 
observation we computed the predicted rms of the mismodeled ionospheric 
contribution as

{\small \beq
  b = A \sqrt{\Cov_{11}^2 \, M_1^2(e) - 2 \Cov_{12} \, M_1(e) \, M_2(e) +
         \Cov_{22}^2\, M_2^2(e) }, \hspace{-0.75em}
\eeq{e:e3} }
  where $M_1(e)$ and $M_2(e)$ are the mapping function of the ionospheric 
path delay. These parameters $b$ were used for weight corrections in
expression~\ref{e:e1}.

  Parameter $A$ varied from 0.35 to 0.59 with a mean of 0.48 for the LCS 
campaign. This means that applying the ionospheric contribution from the 
CODE TEC maps, we reduce the variance of the total contribution by a 
factor of 2, and the mismodeled part of the contribution is accounted in 
inflating uncertainty of group delay. The known deficiency of this approach 
is that first, the regression dependence of parameter on $A$ on GTEC is rather 
coarse, and second, the correlations between residual ionospheric 
contributions are neglected.

  For a check of the contribution of remaining systematic errors, we compared
our source positions derived from X-band only LCS experiments with results
of dual-band observations that included some LCS target sources. In 2017,
the SOuthern Astrometry Program (SOAP) of dual-band follow-up observations at
the Hh-Ho-Ke-Yg-Wa-Ww-Pa network at 2.3/8.4~GHz commenced. The goal of the 
program is to improve the positions of the bright sources with declinations 
below  $-45^\circ$. By August 2018, 10 twenty-hour experiments were observed. 
{\sc parkes} station participated in two of them. The sources as weak as 
70~mJy were observed in experiments with {\sc parkes}, 2--3 scans per sources, 
and objects brighter than 250~mJy were observed in other experiments, 8--10 
scans per source. These experiments were made with the so-called geodetic 
frequency setup: 6 IFs of 16~MHz wide were spanned between 2.200 and 2.304~GHz 
(S-band) and 10 IFs of 16~MHz wide were spanned between 8.198 and 8.950~GHz 
(X-band). Group delays were computed for X and S band separately, and the 
ionosphere-free combinations of group delays were formed. At the moment of 
writing, the program has not finished, and a detailed analysis will be 
presented in the future upon completion of the program. Meanwhile, we use 
these 10 experiments to compare results and assess the errors.

\begin{figure}
    \centerline{\includegraphics[width=0.48\textwidth]{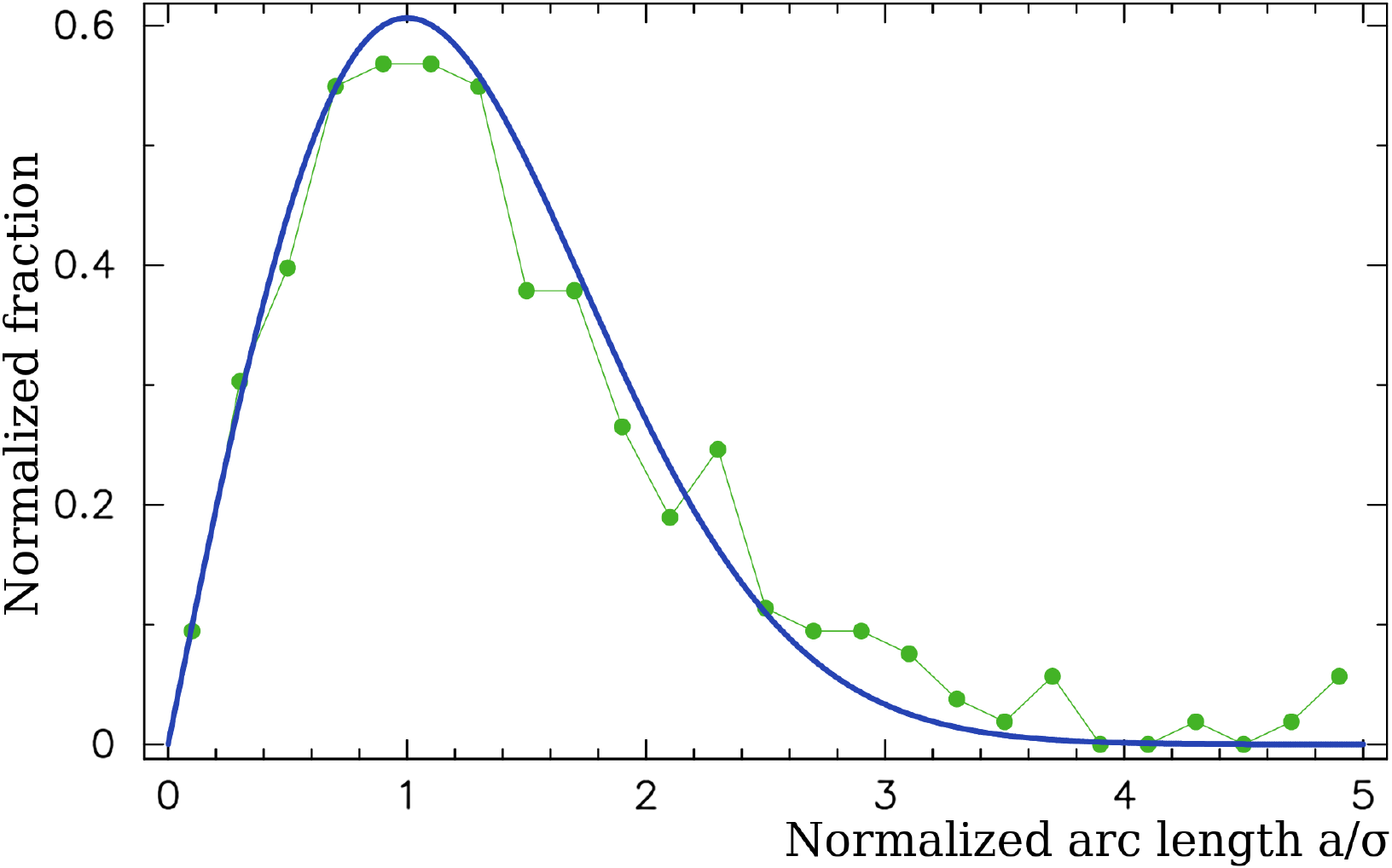}}
    \caption{The distribution of the normalized arc lengths between LCS
             X-band only positions of 269 sources and their X/S positions
             from the follow-up campaigns (Green dots). For comparison,
             the Rayleigh distribution with $\sigma=1$ parameter is shown
             with a blue line.}
    \label{f:rayleigh}
\end{figure}

  We ran a global reference solution using all dual-band X/S observations
of the LCS target sources including the SOAP observations and excluding
LCS observations. The reference and the LCS solutions differed 1) in
the list of sessions that were used in the solutions and 2) in treatment
of the ionosphere. The reference solution used ionosphere-free linear
combinations of S and X-band observables, while the LCS solution used X-band
only group delays with the ionosphere contribution derived from CODE TEC 
maps applied during data reduction. The reference solution used weights 
according to equation \ref{e:e1} with k=1 and b=0.

\begin{figure}
    \centerline{\includegraphics[width=0.44\textwidth]{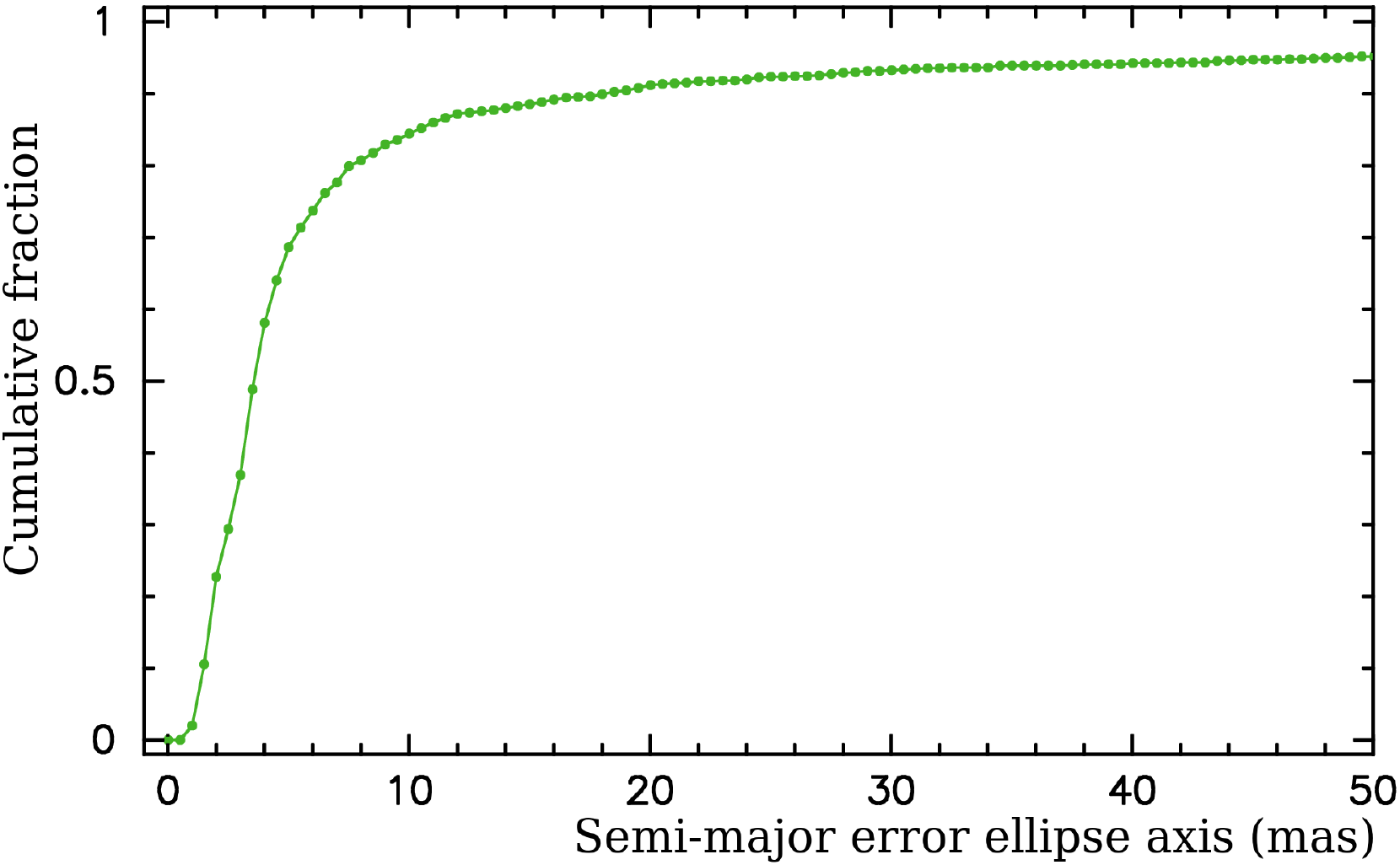}}
    \caption{The cumulative distribution of the LCS position errors.}
    \label{f:lcs2errs}
\end{figure}

   We have compared the positions of 373 LCS target sources that
are common with the reference solution. We did not find any outlier
exceeding 20~mas that can be caused by errors in group delay ambiguity
resolution. That means that all observations with unreliable ambiguity
resolution were correctly flagged out and did not degrade the solution.
At the same time, we found that the arc lengths divided by the their
uncertainties, so-called normalized arcs, were larger than expected with
the mean value 1.89. We attributed this discrepancy to the
underestimation of errors of LCS observations. To alleviate this
underestimation, we varied the multiplicative factor $k$ in expression
\ref{e:e1} in such a way the distribution of normalized arcs be as close
to the Rayleigh distribution with $\sigma=1$ as possible. We found that 
when the LCS observations are re-weighted with parameter $k=1.80$ in 
equation \ref{e:e1}, the distribution of normalized arcs over 269~sources
that have at least 16 observations is the closest to the Rayleigh 
distribution (see Figure~\ref{f:rayleigh}). The mean arc length is 
3.4~mas and the median value is 2.5~mas. The cumulative distribution of 
the final LCS position errors is shown in Figure~\ref{f:lcs2errs}.

\section{The catalogue}

  The first 8 rows of the LCS catalogue are presented in Table~\ref{t:lcscat}.
The catalogue presents source positions, position uncertainties, the
number of used observations, flux densities in three ranges of baseline
projection lengths, and their formal uncertainties. In total, the catalogue
has 1100 entries. The median semi-major error ellipse axes of reported
positions is 3.6~mas. The flux densities are in a range from 3~mJy to 2.5~Jy,
with the median 102~mJy. For completeness, the list of 405 sources that have
been observed, but not detected is given in the machine-readable table
datafile3. The flux densities of such sources turned out to be below the 
detection limit of baselines 
{\sc parkes}/{\sc atca}, {\sc parkes}/{\sc hobart26}, {\sc parkes}/{\sc ceduna} 
that is typically 6--8~mJy.

\begin{table*}
   \hspace{-2.0em}
   \scalebox{0.88}{
      \begin{tabular}{rrrrrrrrrrrrrr}
          \hline
            \ntab{c}{(1)}  & \ntab{c}{(2)}  & \ntab{c}{(3)}  & \ntab{c}{(4)} & \ntab{c}{(5)}  &
            \ntab{c}{(6)}  & \ntab{c}{(7)}  & \ntab{c}{(8)}  & \ntab{c}{(9)} & \ntab{c}{(10)} &
            \ntab{c}{(11)} & \ntab{c}{(12)} & \ntab{c}{(13)} & \ntab{c}{(14)} 
                \\
                           &   & hh mm ss.fffff \hphantom{.} & $ {}^\circ \hspace{1em} {}' \hspace{1em} {}'' \hphantom{aaa}$ 
                           & mas & mas & & & Jy & Jy & Jy & Jy & Jy & Jy \\
          \hline
          LCS J0001$-$4155 & 2358$-$422 & 00 01 32.75494 & $-$41 55 25.3367 & 215.1 & 92.3 & -0.904 &    5 &  0.008 &  0.007 & -1.0   &  0.001 & 0.002 & -1.0   \\
          LCS J0002$-$6726 & 2359$-$677 & 00 02 15.19280 & $-$67 26 53.4337 &  89.6 & 32.8 &  0.553 &    5 &  0.006 &  0.007 & -1.0   &  0.001 & 0.001 & -1.0   \\
          LCS J0002$-$5621 & 0000$-$566 & 00 02 53.46830 & $-$56 21 10.7831 &  23.8 &  9.3 &  0.421 &    8 &  0.172 &  0.052 &  0.141 &  0.013 & 0.012 &  0.031 \\
          LCS J0003$-$5444 & 0000$-$550 & 00 03 10.63084 & $-$54 44 55.9923 &  42.1 & 10.7 & -0.112 &    9 &  0.006 &  0.006 & -1.0   &  0.001 & 0.001 & -1.0   \\
          LCS J0003$-$5247 & 0000$-$530 & 00 03 19.60042 & $-$52 47 27.2834 &  39.0 & 18.5 & -0.291 &    8 &  0.013 &  0.014 & -1.0   &  0.002 & 0.002 & -1.0   \\
          LCS J0004$-$4345 & 0001$-$440 & 00 04 07.25762 & $-$43 45 10.1469 &   4.0 &  3.0 &  0.163 &   44 &  0.188 &  0.205 &  0.214 &  0.030 & 0.024 &  0.046 \\
          LCS J0004$-$5254 & 0001$-$531 & 00 04 14.01314 & $-$52 54 58.7099 &   8.8 &  3.7 &  0.039 &   36 &  0.027 &  0.027 &  0.018 &  0.002 & 0.003 &  0.003 \\
          \hline
      \end{tabular}
   }\hfill
   \caption{The first 7 rows of the LCS catalogue. Columns: (1) source ID;
            (2) alternative source name; (3) J2000 right ascension;
            (4) J2000 declination; (5) uncertainty in right ascension without
            $\cos\delta$ factor;
            (6) uncertainty in declination; (7) correlation between right
            ascension and declination estimates; (8) the number of
            observations used in the solution; (9) the median correlated
            flux density \note{at 8.3~GHz} at baseline projection lengths 
            in a range 0--10~M$\lambda$; (10) the median correlated flux 
            density at baseline projection lengths in a range 10--40~M$\lambda$;
            (11) the median correlated flux density at baseline projection
            lengths in a range 40--300~M$\lambda$; 12) the median uncertainty
            of the correlated flux density at baseline projection lengths
            in a range 0--10~M$\lambda$; (13) the median uncertainty of the
            correlated flux density at baseline projection lengths in a range
            10--40~M$\lambda$; (14) the median uncertainty of the correlated
            flux density at baseline projection lengths in a range
            40--300~M$\lambda$.
            This table is available in its entirety in machine-readable
            table datafile1 and Virtual Observatory (VO) forms in the online
            journal. A portion is shown here for guidance regarding its form
            and content. }
    \label{t:lcscat}
\end{table*}

\begin{figure}
    \includegraphics[width=0.48\textwidth]{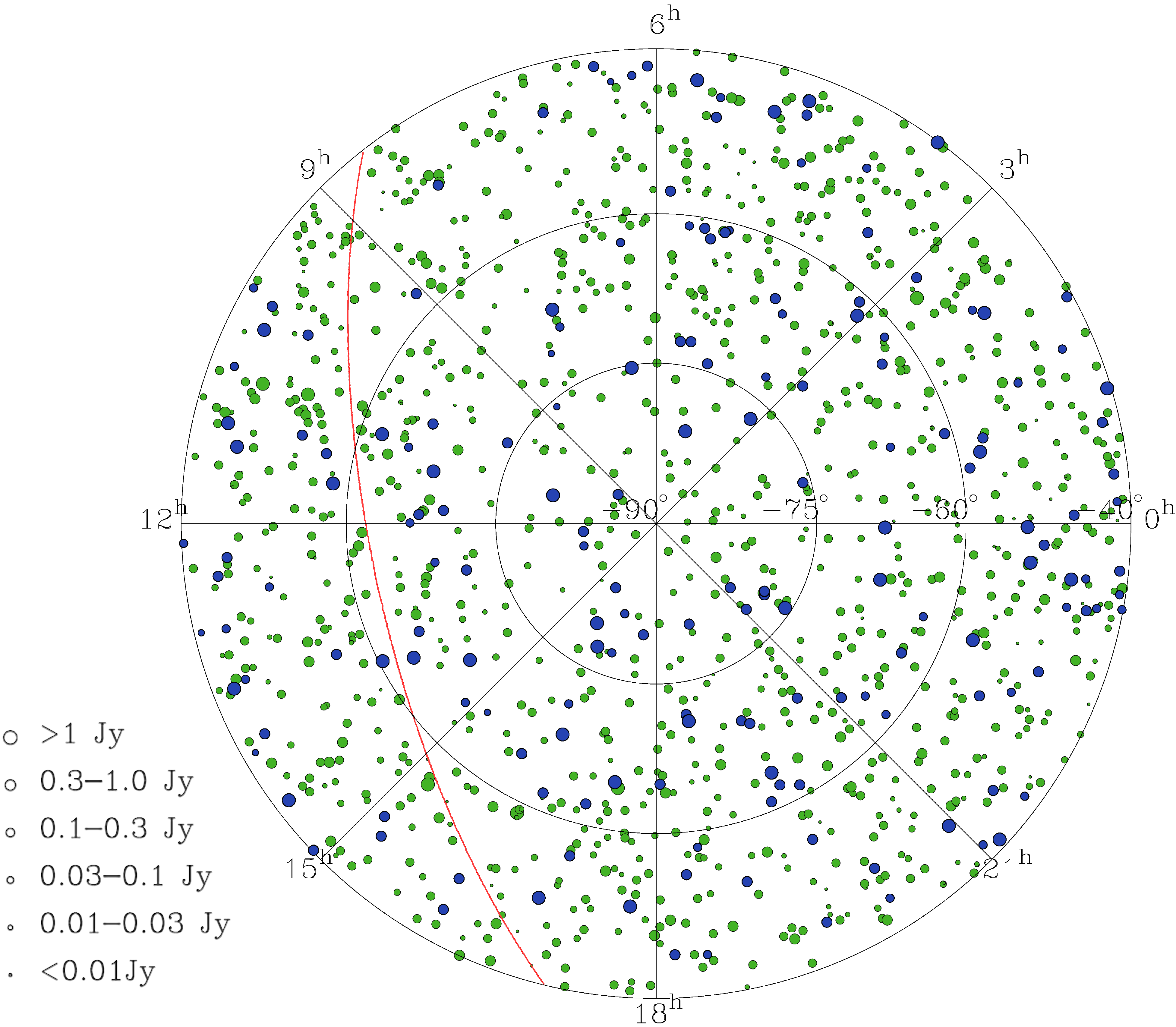}
    \caption{The sky distribution of compact radio sources at the Southern
             Hemisphere \note{at 8.3~GHz}. Blue light color denotes 186 sources 
             with declinations $< -40^\circ$ with VLBI positions known prior
             the LCS program. Green dark color denotes 1100 sources
             detected in LCS program. Red line shows the Galactic plane.}
    \label{f:skydist}
\end{figure}

  The distribution of LCS sources on the sky is shown in
Figure~\ref{f:skydist}. The distribution is rather uniform and does not have
avoidance zones. For comparison, the sources known priori the LCS
campaign are shown with blue color.

\section{Discussion}

  The median position uncertainty, 3.6~mas, cannot be called the VLBI state
of the art nowadays. There are four factors that played the role. First,
the contribution of the ionosphere cannot be computed using GNSS TEC models
with the same level of accuracy as using simultaneous dual-band observations.
Second, the scale of the network, less than 1700 km for the most observations
degraded the sensitivity of observations to source positions, since the
source position uncertainty is reciprocal to the baseline length. Station
{\sc hartrao} participated in less than 25\% observations due to 
scheduling constraints. Third, the spanned bandwidth was limited to 320~MHz, 
compare with 720~MHz typically used in geodetic VLBI. Positional uncertainty 
is approximately reciprocal to the spanned bandwidth. Fourth, observed 
sources were rather week: 25\% of target sources are weaker than 46~mJy.

\begin{figure}
    \includegraphics[width=0.48\textwidth]{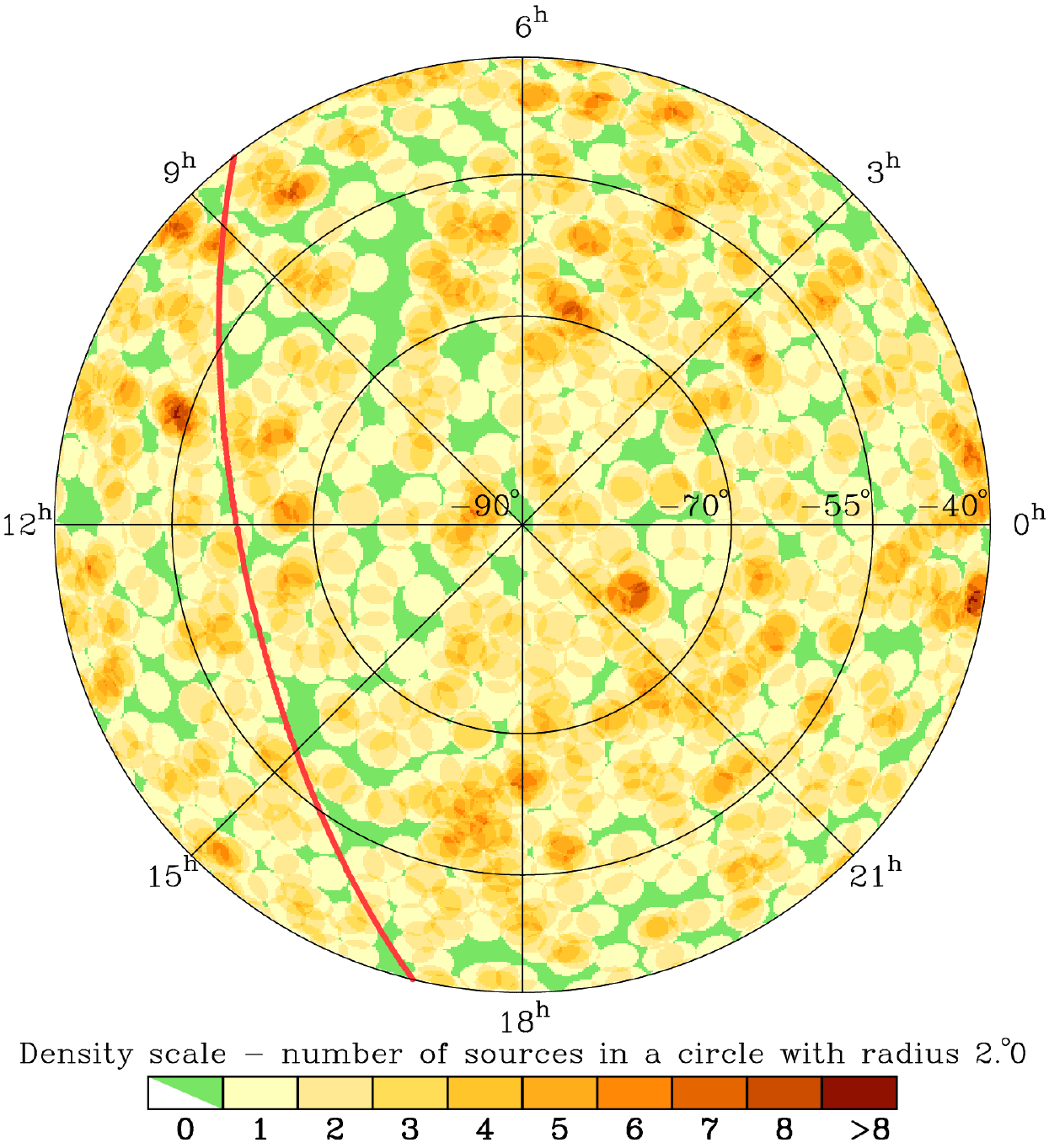}
    \caption{The sky density of calibrator sources \note{at 8.3~GHz} in the zone 
             with declinations $< -40^\circ$ defined as the number of 
             compact sources with flux density $> 30$ mJy in a circle 
             of $2^\circ$ radius. The Galactic plane is shown with 
             the red line.}
    \label{f:caldens}
\end{figure}

  Nevertheless, position accuracy of several milliarcseconds is sufficient
for phase referencing. Figure~\ref{f:caldens} shows the probability of
finding a phase calibrator brighter than 30~mJy within $2^\circ$ of any 
target with $\delta < -40^\circ$. For 88\% of the area, such a calibrator 
can be found. Information about flux densities of detected sources and
the upper limits for undetected sources greatly facilitates future follow-up
observing programs focused on improvement of source positions.

   Among the 1100 LCS sources, there are 725 counterparts with \Gaia\ DR2
\citep{r:gaia_dr2} with the probability of false detection below 0.0002.
See \citet{r:gaia1} for detail of the VLBI and \Gaia\ association procedure.
\citet{r:gaia4} showed that comparison of over 9,000 matched VLBI/\Gaia\
sources reveled that 9\% have statistically significant offsets at
the level exceeding $4\sigma$. They presented extensive arguments showing that
these offsets are real and are a manifestation of the presence of optical
jets that affect the positions of optic centroid reported by \Gaia. The LCS
dataset has 53 (7.2\%) outliers with an arc length exceeding $4\sigma$. The
lower fraction of outliers is explained by worse position accuracy. These
outliers were excluded from further analysis, since they do not characterize
catalogue errors. The median arc length of position differences is
3.2~mas, while the median semi-major error ellipse axes of LCS positions of
matched sources is 3.3~mas and the median semi-major error ellipse of \Gaia\
positions of matched sources is 0.3~mas. This comparison demonstrates that
the median of the differences between LCS and \Gaia\ positions is very
close to the reported median of the LCS semi-major error axis.

\begin{figure}
    \includegraphics[width=0.46\textwidth]{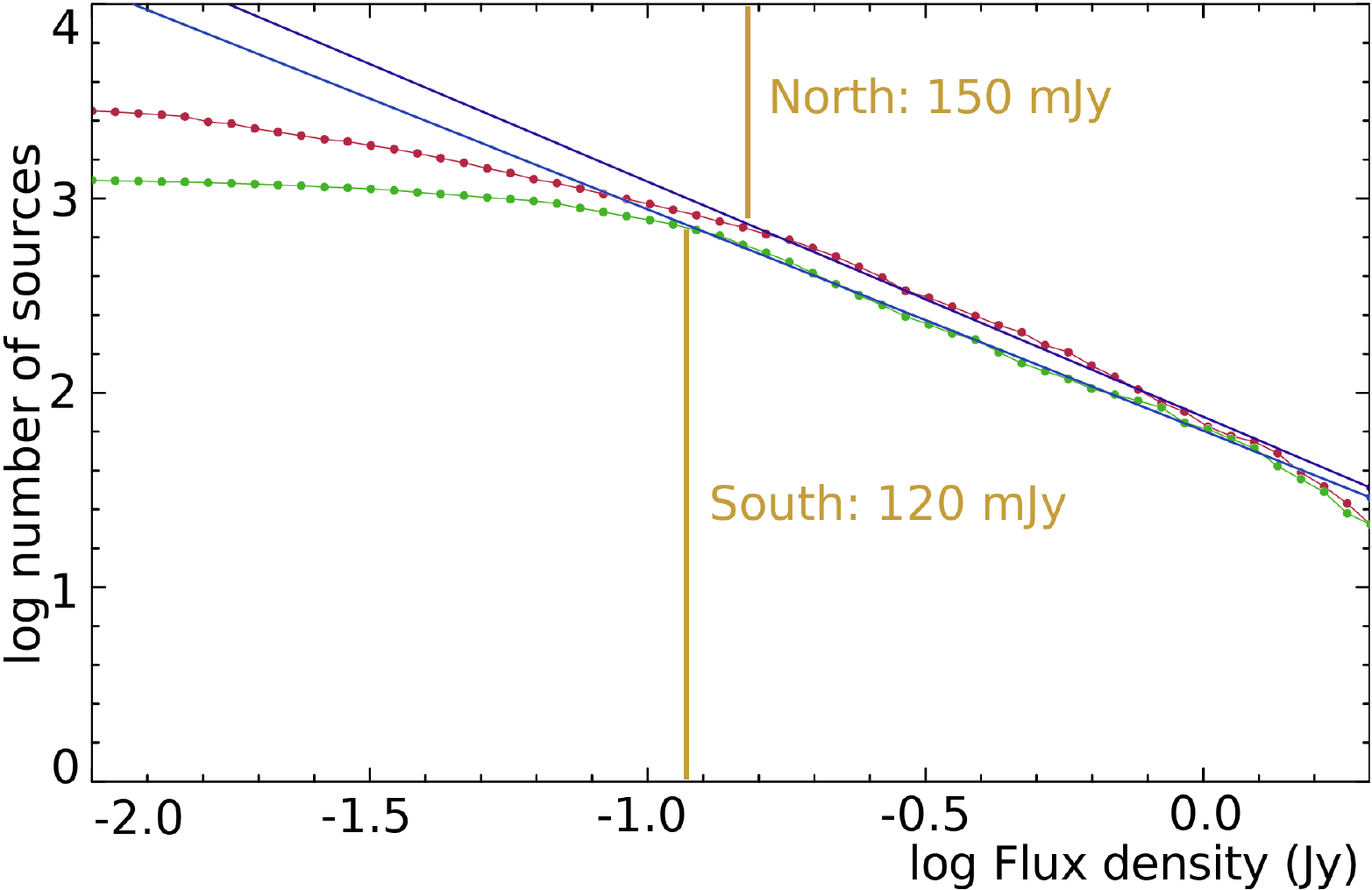}
    \caption{The log N -- log S diagram for the LCS catalogue (low green
             line) using only the sources with $\delta < -40^\circ$.
             The upper red line shows a similar diagram for the
             sources with $\delta > +40^\circ$ at 8~GHz observed under 
             other programs.}
    \label{f:lognlogs}
\end{figure}

  For analysis of LCS completeness we computed the so-called log N -- log S
diagram --- the dependence of the logarithm of the number of sources
on the logarithm of the total flux density recovered from VLBI observations.
The dependence is approximated by a straight line within the range
of flux densities that the catalogue is considered complete. With a decrease
of flux densities, at some point the diagram deviates from a straight
line. This point is considered the limit below which the catalogue
is incomplete. The diagram in Figure~\ref{f:lognlogs} shows the completeness
level of the LCS subsample at $\delta < -40^\circ$ drops below 95\% at
flux densities 120~mJy.

  For comparison, we computed a similar diagram for the Northern Hemisphere
at declinations $>+40^\circ$ using the Radio Fundamental Catalogue. The
Northern Hemisphere catalogue has more weak sources, but surprisingly, its
completeness drops below 95\% level at flux densities 150~mJy. At the same
time, the Northern Hemisphere catalogue has 23\% more sources.
One explanation is the total number of sources in the Southern Hemisphere
is indeed $\sim\! 20$\% less due to a large scale fluctuation of the source
distribution over the sky. Another explanation is a selection bias.
The parent catalogue of the LCS is the AT20G at 20~GHz, while the parent
catalogues of the Northern Hemisphere sources were observed at lower
frequencies: 5--8~GHz. Selecting sources based on their emission at 20~GHz
may result in omitting the objects with falling spectrum. 
The \mbox{log N -- log S} dependencies for southern and Northern Hemispheres 
are almost parallel in a range of 0.15--0.65~Jy. If we accept the hypothesis 
that selecting candidate sources based on AT20G catalogue causes a bias, 
we have to admit that using AT20G as a parent sample we lose sources as 
bright as 0.5~Jy, which is difficult to explain. We think the problem of 
completeness of the LCS is still open, and more observations are needed 
in order to resolve it.

\section{Compact and extended emission in a sub-sample of compact 
         sources from AT20G}

  Of a sub-sample of 907 AT20G sources observed in the LCS campaign, 
839 or 93\% were detected. As noted by \cite{r:chh13}, most of these 
AT20G sources are already known to be compact on scales of 
$\sim0.15''$ at high frequencies.

\begin{figure}
    \ifpre 
        \centerline{\includegraphics[width=0.37\textwidth]{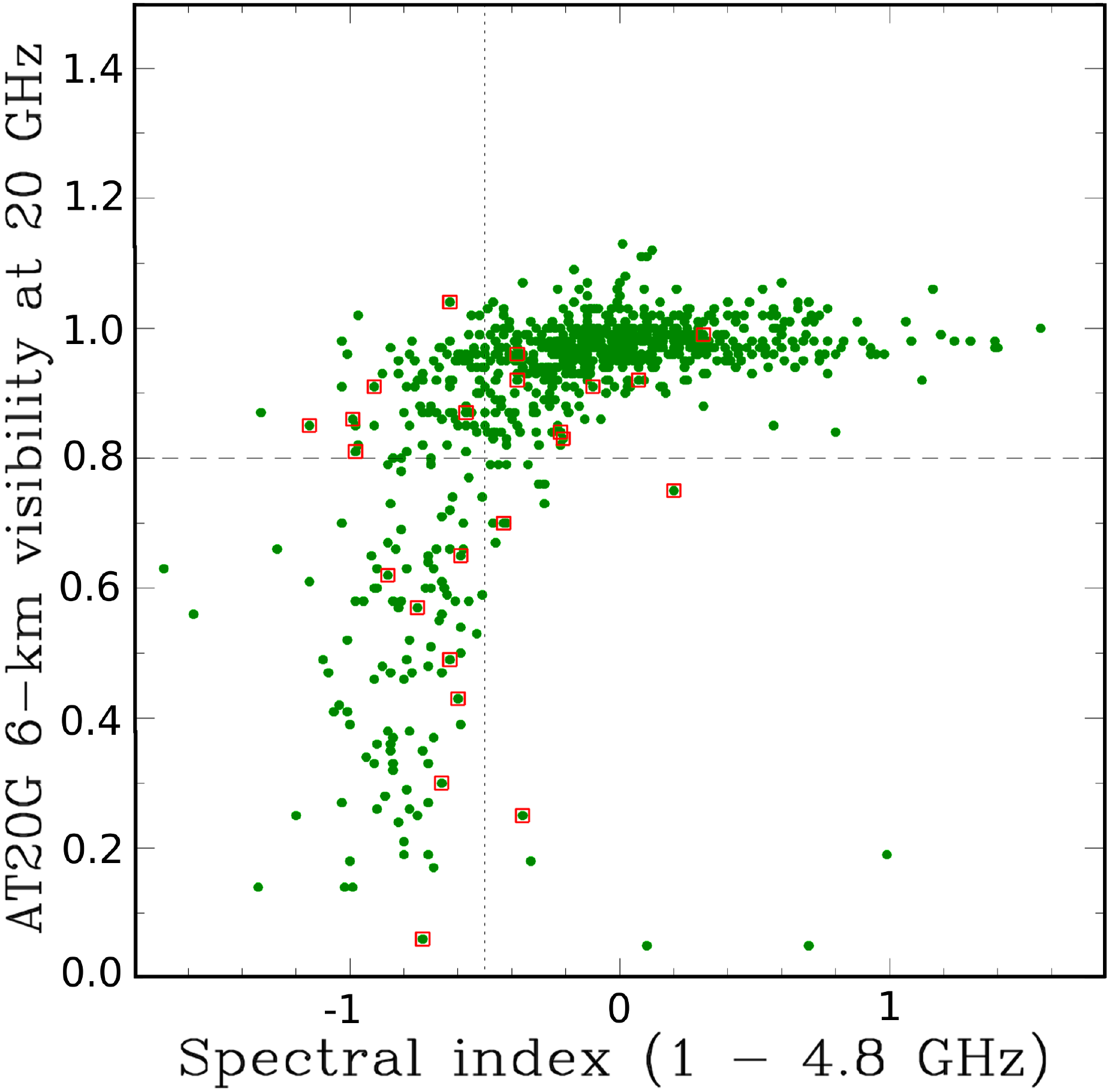}}
    \else
        \centerline{\includegraphics[width=0.45\textwidth]{lcs2_popan_01.pdf}}
    \fi
    \caption{The ratio of the flux density at the 6~km long ATCA baseline 
             to the flux density at short baselines from AT20G observations 
             at 20~GHz (adapted from Figure~7 of \citet{r:chh13}).
            }
    \label{f:at20g_6km}
\end{figure}

  Figure \ref{f:at20g_6km} plots the 6-km visibilities (defined as the  
ratio of the 20\,GHz flux density on 6~km long baselines to the flux 
density at short baselines of 30--60~m; see \citet{r:chh13} for details) 
for the subsample of AT20G sources observed in the LCS campaign that 
also have 8.4\,GHz flux density measurements in the AT20G catalogue. 
The red squares in this figure show local AT20G galaxies from the study 
by \citet{r:sad14}. Compact flat-spectrum AGN lie in the upper right 
quadrant and compact steep-spectrum (CSS) sources in the upper left, 
while the sources in the lower left quadrant are radio AGN with 
extended emission (mainly FR-1 radio galaxies) and angular sizes 
larger than 150~mas.

\begin{figure}
    \ifpre 
        \centerline{\includegraphics[width=0.37\textwidth]{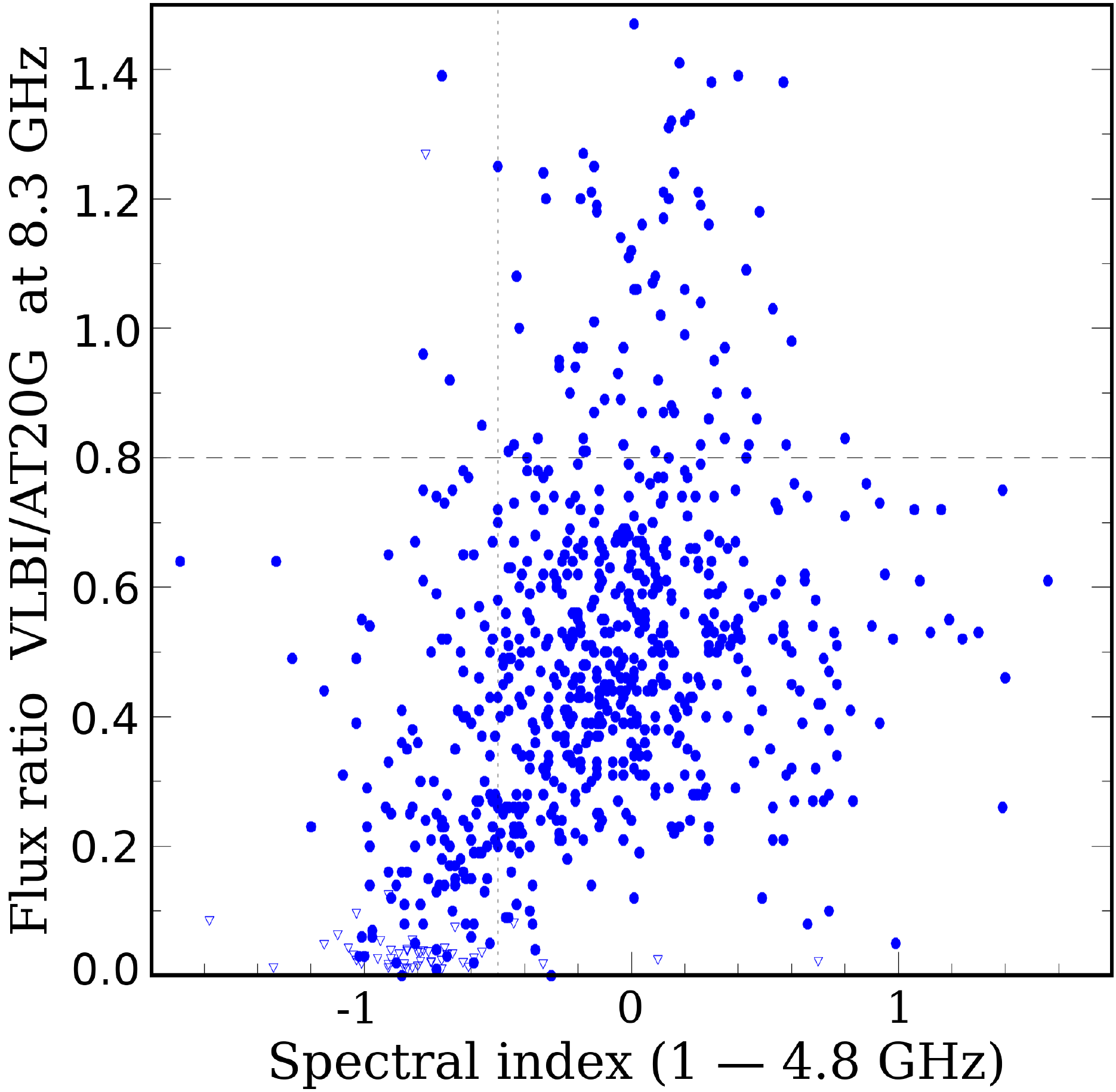}}
    \else
        \centerline{\includegraphics[width=0.45\textwidth]{lcs2_popan_02.pdf}}
    \fi
    \caption{The ratio of the flux density from LCS at 8.3~GHz to the 
             total flux density at 8.4~GHz from AT20G observations.
             The triangles show upper limits for the sources that were 
             undetected on VLBI baselines.}
    \label{f:at20g_vlbi}
\end{figure}

  Figure~\ref{f:at20g_vlbi} shows a similar plot where the compactness on 
much smaller scales is estimated by using the ratio of the  
flux density derived from VLBI observations at 8.3~GHz to the total flux 
density at 8.4~GHz from the AT20G observations. The majority  of the 
flat-spectrum AT20G sources, while still detected on VLBI baselines, are 
now resolved at 10--40~M$\lambda$, which corresponds to 5--20~mas angular 
size. The sources with spectral index steeper than 
-0.5 show systematically lower compactness than the sources with flat radio 
spectra. The majority of VLBI non-detections (65 out of 68) are AT20G 
sources with steep radio spectra.

  The sub-sample of AT20G sources observed in the LCS is not complete, 
and further analysis is outside the scope of this paper. The compactness 
plots are shown here to demonstrate the potential of the LCS dataset, 
and a detailed analysis will be presented in a future paper. 

\section{Summary}

   The LCS VLBI observing program has provided positions of 1100 compact 
radio sources at declinations below $-30^\circ$ with accuracies at
a milliarcsecond level and estimates of flux density at 8.3~GHz.
As a result, the number of compact radio sources south of declination
$-40\degr$ which have measured VLBI correlated flux densities and positions
known to milliarcsecond accuracy has increased by a factor of 6.4.
\Note{A dense grid of calibrator sources with precisely known positions is
important for a number of applications, in particular for support of ALMA
observations.} The internal LCS test based on the \mbox{log N -- log S} 
diagram shows it is complete at a 95\% level for sources brighter than 
120~mJy. At the same time, comparing the LCS with the Northern Hemisphere 
catalogue, we found a $\sim\! 20$\% difference in the source count. 
The LCS may have a deficiency of $\sim\! 20$\% sources 
because of using AT20G as a parent sample. It is yet to be resolved 
whether using high-frequency parent catalogue results in a systematic 
loss of sources with falling spectrum. The LCS catalogue is the Southern 
Hemisphere counterpart of the VLBA Calibrator Survey. The major outcome 
of this campaign is elimination of the hemisphere bias that the VLBI 
catalogues suffered in the past. However, technical limitations of the 
Southern Hemisphere telescopes provided position accuracy one order of 
magnitude worse than the accuracy of similar catalogues in the Northern 
Hemisphere. Future observations will target LCS sources for improvement 
of their positions, and the first such follow-up observing campaign 
started in 2017.

\section*{Acknowledgments}

  The authors would like to thank Jonathan Quick, Jamie Stevens, 
Jamie McCallum, Lucia McCallum, Jim Lovell, Sergei Gulyaev, Tim Natusch, 
Stuart Weston, Phillip Edwards, Cormac Reynolds, Michael Bietenholz, 
and Alessandra Bertrarini  for help in observations and correlation. 

  The Long Baseline Array is part of the Australia
Telescope National Facility which is funded by the Commonwealth of
Australia for operation as a National Facility managed by CSIRO.
This work was supported by resources provided by the Pawsey Supercomputing 
Centre. The Australian SKA Pathfinder is part of the Australia  Telescope 
National Facility which is managed by CSIRO. Operation of ASKAP is funded 
by the Australian Government with support from the National Collaborative 
Research Infrastructure Strategy.  Establishment of ASKAP, the Murchison 
Radio-astronomy Observatory and the Pawsey Supercomputing Centre are 
initiatives of the Australian Government, with support from the Government 
of Western Australia and the Science and Industry Endowment Fund. 
We acknowledge the Wajarri Yamatji people as the traditional owners of the 
ASKAP observatory site.

\label{lastpage}

\bibliographystyle{mnras}
\bibliography{lcs2}

\end{document}